# Differences in Online Course Usage and IP Geolocation Bias by Local Economic Profile

by

Daniela Ida Ganelin

Submitted to the Department of Electrical Engineering and Computer Science
on May 23, 2019 in Partial Fulfillment of the Requirements for the
Degree of Master of Engineering in Electrical Engineering and Computer Science

ABSTRACT

Although Massive Online Open Courses (MOOCs) have the promise to make rigorous higher education accessible to everyone, prior research has shown that registrants tend to come from backgrounds of higher socioeconomic status. In this work, I study geographically granular economic patterns in registration for HarvardX and MITx courses, and in the accuracy of identifying users' locations from their IP addresses. Using ZIP Codes identified by the MaxMind IP geolocation database, I find that per-capita registration rates correlate with economic prosperity and population density. Comparing these ZIP Codes with user-provided mailing addresses, I find evidence of bias in MaxMind geolocation: it makes greater errors, both geographically and economically, for users from more economically distressed areas; it disproportionately geolocates users to prosperous areas; and it underestimates the regressive pattern in MOOC registration. Similar economic biases may affect IP geolocation in other academic, commercial, and legal contexts.

Thesis Supervisor: Isaac Chuang
Title:   Professor of Electrical Engineering and Computer Science
         Professor of Physics
         Senior Associate Dean of Digital Learning

# Table of Contents

# 1. List of figures

# 2. List of tables

# 3. Acknowledgements

Thanks to Isaac Chuang for his wise, compassionate, and inquisitive mentorship.

To Mohammad Alizadeh, Hari Balakrishnan, Andrew Ho, and Joshua Goodman for feedback on this work, advice, and interesting classes.

To MIT Open Learning and Sanjay Sarma for financial support, edX and the Economic Innovation Group for data, and the Harvard-MIT Data Center for computational resources.

To Kristina Heavey, Anne Hunter, Stefanie Jegelka, and Katrina LaCurts for life advice and extraordinary administrative flexibility.

To GTL Wales and Liam and Carolyn Rahman for a wonderful, much-needed midyear reset, and to Concert Choir for weekly ones.

To Cambridge's shockingly many coffee shops for caffeine and comfortable chairs.

To French House, Jakob Coray, and Anastasia Tyurina for providing warm surrogate homes, and both emotional and culinary support.

Finally, to my parents and the rest of my incredible family, for everything.

# 4. Introduction

In this introduction, I explain the motivation for the two parts of my thesis (4.1) and outline my research questions (4.2), approach (4.3), and key results (4.4). Chapter 5 contains more detailed background.

## 4.1. Motivation

### 4.1.1. Tracking local economic patterns in MOOC usage

In 2012, MIT and Harvard launched edX, a platform that allows users around the world to participate in free massive open online courses (MOOCs) (MIT Open Learning 2015). Millions of users have since enrolled in hundreds of courses on either HarvardX or MITx. Many registrants never open or engage with course material; some work through part of a course; a small fraction complete courses and earn certificates (Chuang and Ho 2016), which may involve paying fees for ID verification (Hashmi 2013).

When Massive Online Open Courses (MOOCs) first arose, many saw them as a means of fundamentally altering systems of higher education and eliminating economic inequities. edX CEO Anant Agarwal said that "We hope to democratize and reimagine education so that anyone, anywhere, regardless of his or her social status or income, can access education" (Kanani 2014). The presidents of Harvard and MIT described "the power to democratize access to education on a scale never seen in history" (Faust and Reif 2013). New York Times columnist Thomas

Friedman wrote that "nothing has more potential to lift more people out of poverty... Nothing has more potential to unlock a billion more brains to solve the world's biggest problems" (Friedman 2013).

Within the United States, there is wide and growing economic disparity across geographic areas (Shambaugh and Nunn 2018), even within cities (Misra 2018). MOOC providers might hope that people in less prosperous regions preferentially take advantage of MOOCs to replace or supplement traditional, expensive higher education or to boost their employability. Given evidence to the contrary, providers might aim to improve relevance, marketing, and accessibility of courses in these areas.

Studying economic patterns in MOOC usage is difficult because researchers generally have no direct measures of users' socioeconomic backgrounds: at best, some users self-report their level of education. One workaround is to approximate users' socioeconomic status by identifying their locations, from self-reported addresses or their IP addresses, and using socioeconomic properties of the geographic area rather than of the individual. The same logic underlies the new neighborhood-based SAT "adversity index" for college applicants (Jaschik 2019).

Both of these steps are error-prone. First, the identified location may be incorrect (explained in 5.4 and 8.4). Second, the area may be a poor proxy for the individual; research in other fields has found mixed evidence (5.6).

A handful of studies have examined economic patterns in MOOC usage (reviewed in 5.1). They have consistently found that MOOC users tend to come from individual and geographic backgrounds of higher socioeconomic status. These results raise doubts about whether MOOCs are meeting their key goals: are they providing access to education, and so economic advancement, to people who could not otherwise afford it?

Only one study has focused on U.S. MOOC users' economic backgrounds. Hansen and Reich 2015 found that edX were disproportionately located in wealthy, well-educated neighborhoods. They used self-reported mailing addresses from a fraction of users. No prior study has examined a large, representative sample of MOOC users.

### 4.1.2. Assessing accuracy and bias of IP geolocation

Some MOOC research, including my own, identifies users' location by IP geolocation: using a commercial dataset to translate users' recorded Internet Protocol (IP) addresses to approximate locations. IP geolocation is appealing to researchers as an easily scalable, inexpensive tool. It is also used in other social science research, computer science research, and many commercial and governmental applications.

Unfortunately, studies have consistently found that IP geolocation is often substantially inaccurate, particularly at granular levels (overviewed in 5.5). Errors can occur because the recorded IP addresses do not correspond to users' physical locations or because the databases are inaccurate. MOOC papers that use the technology seldom mention concerns about its accuracy (5.2), and to the best of my knowledge no study has examined the accuracy of IP geolocation specifically in a MOOC context.

Particularly when used to study economic patterns, there is the possibility that IP geolocation is not just frequently erroneous, but also systematically biased - for example, it may be more accurate in places with greater prosperity and population density, where Internet infrastructure is more developed (as discussed in Whitacre and Mills 2007). I am familiar with no previous research on such biases.

## 4.2. Research questions

This thesis is divided into two main parts. The first concerns the economics of MOOC usage, using geolocated ZIP Codes under the assumption that IP geolocation is accurate. The second challenges this assumption and focuses on the accuracy of IP geolocation, using MOOC data as a tool for measuring geolocation rather than as the primary object of study.

In the first part (Chapter 7), I study the differences in edX usage by ZIP-Code level economic properties, identifying users' locations with IP geolocation. This work builds on Hansen and Reich by using a more complete user population, a more familiar geographic unit, and a broader set of economic indicators. My research questions are:

- What is the distribution of per-capita rates of course registration, completion, and certification over U.S. ZIP Codes?

- What are the relationships between ZIP Code course engagement rates and ZIP Code economic properties?
- Are ZIP Code course engagement levels unidimensional? If so, could this trait constitute a "MOOC Propensity" which is more helpful than simple registration counts in predicting economic factors? If not, what predicts course-by-ZIP enrollments?

In the second part (Chapter 8), I assess the accuracy and bias of IP geolocation using "ground-truth" ZIP Codes extracted from mailing addresses provided by a portion of users. I ask:
- According to ground-truth data, how do per-capita registration rates vary with ZIP Code economic properties?
- How do the distributions of geolocation error vary with ground-truth ZIP Code economic properties?
- How do the economic profiles of IP-geolocated ZIP Codes vary with the economic properties of ground-truth ZIP Codes?
- How do the overall distributions of users' economic backgrounds differ between ground-truth and IP-geolocated data?

## 4.3. Approach

In Chapter 7, I apply psychometric methods to study ZIP Codes' course usage, modelling MOOC usage as an exam in which each ZIP Code is a test-taker and each course is an exam question. I use simple descriptive statistics and maps to find general patterns in the data and examine relationships between ZIP Code usage and ZIP Code economic properties. I apply classical test theory to check that this exam is reliable (internally consistent). I use factor analysis to establish unidimensionality in the data, and correlational techniques to understand the outputs of factor analysis.

In Chapter 8, I measure the bias in IP geolocation on this dataset. I extract "ground-truth" ZIP Codes from user-provided mailing addresses, and compare them to ZIP Codes found by MaxMind from IP addresses. I calculate the geographic distance between them, and examine the distributions of this distance based on the properties (prosperity, area, population, and density) of the ground-truth ZIP Code. I also examine the distribution of IP-geolocated ZIP Code properties by ground-truth ZIP Code properties. I use linear regression to study multivariate interactions among ZIP Code properties, geolocation accuracy, and ZIP Code identifications.

## 4.4. Summary of key results

In Chapter 7, I find that, according to IP-geolocated ZIP Codes, edX users are more likely to come from economically prosperous, well-educated, high-density areas, substantiating Hansen and Reich's results over a nearly complete user population. Registration rates across different courses are largely unidimensional: considering per-course registrations provides little more information than total counts. Different courses behave much alike in terms of their geographic distribution.

In Chapter 8, I find that according to locations derived from mailing addresses, too, edX users tend to come from prosperous, dense ZIP Codes. However, I also find that IP geolocation is biased in several ways: it makes greater errors in ZIP Codes with lower prosperity or population density; it misidentifies users' economic backgrounds more often for users from less prosperous locations; and it underestimates the gap in MOOC usage between prosperous and distressed places. To the best of my knowledge, this is the first work on economic bias introduced by IP geolocation.

# 5. Literature review

In this chapter, I provide an overview of the literature on economic patterns in MOOC usage (5.1), and of past MOOC research that has used IP geolocation (5.2). Changing focus to IP geolocation, I describe a sampling of its other applications (5.3), summarize its technical workings (5.4), and overview the networking literature on its effectiveness (5.5). I also briefly review literature on geographic proxies for socioeconomic status (5.6).

## 5.1. The economics of MOOC usage

Studies have consistently found that MOOC participants tend to come from individual and geographic backgrounds of higher socioeconomic status.

Some of these analyses focus on per-country usage, identifying users' countries by IP geolocation. Among registrants for the University of Pennsylvania's MOOCs on Coursera, Christensen et al. report that the large majority are from developed countries, particularly the United States (Christensen et al. 2013). Studying HarvardX and MITx, Chuang and Ho find a strong, exponential relationship between a country's Human Development Index (HDI) and the number of certificates earned by its residents (adjusting for population) (Chuang and Ho 2016). Kizilcec et al. find a similar link between HDI and completion rate among Coursera registrants (Kizilcec et al. 2017; Kizilcec, Piech, and Schneider 2013).

MOOC users across the globe are also highly educated: 73% of HarvardX/MITx survey respondents (Chuang and Ho 2016) and 79% of UPenn respondents (Christensen et al. 2013) report holding a bachelor's degree. Particularly in developing countries, UPenn MOOC users are far more likely to hold college degrees than their countrymates (Emanuel 2013).

Other work focuses on the distribution of users' backgrounds within countries. Again studying UPenn Coursera users, (Alcorn, Christensen, and Kapur 2015) use IP geolocation to estimate that most Indian users come from a handful of "the most populous, developed, and prosperous Indian cities", which account for only a small fraction of India's total population.

Perhaps most relevant to this thesis is Hansen and Reich's research on HarvardX and MITx users (Hansen and Reich 2015b, 2015a). Using some of the same self-reported mailing addresses that I use, the authors match users to census blocks, small neighborhoods that are generally smaller than ZIP Codes. Compared to the general U.S. population, MOOC users come from wealthier, more educated, more densely populated neighborhoods. In addition, registrants who report a higher level of education (or, for adolescents, parental level of education) are more likely to complete courses.

There is one study with good news for users of lower socioeconomic backgrounds. Zhenghao et. al. survey users who completed Coursera MOOCs (Zhenghao et al. 2015). In general, completers with greater socioeconomic status and education are more likely to report that courses benefited their careers - but in developing countries, completers with *lower* socioeconomic status and education report more tangible benefits such as new jobs or raises.

## 5.2. IP geolocation in MOOC research

MOOC studies without an economic focus have also used IP geolocation, to place users to a continent (Kizilcec and Halawa 2015), country (Martinez 2014; Liu et al. 2016; Northcutt, Ho, and Chuang 2015; Guo and Reinecke 2014;
, and sometimes city (Alcorn, Christensen, and Kapur 2015; Breslow et al. 2013) or U.S. ZIP Code (Diver and Martinez 2015).

Some authors do not mention which geolocation database they use, while others use some version of MaxMind, as I do.

These papers seldom raise concerns about the accuracy of the geolocation, instead using strong statements such as "we are able to determine a student's zip code from the IP address information" (Diver and Martinez 2015).

Breslow et al. do note that "there is some error associated with this procedure, as students could log in from proxy servers or otherwise mask their IP address" (Breslow et al. 2013), without mentioning the possibility that the database itself could contain inaccuracies. They also state that "less than 5% of the students were likely to be misidentified due to altered IP addresses", but do not explain how they reach this conclusion (Breslow et al. 2013).

Christensen et. al. compare the distributions of countries according to survey reports and to IP geolocation, aiming to check the representativeness of their survey sample rather than the accuracy of the geolocation. They find "that there was never more than a 1.52% difference between the measured student population according to the survey and the estimated student population according to IP address" (Christensen et al. 2013).

## 5.3. Other applications of IP geolocation

Aside from MOOC studies, IP geolocation is used in other social science and interdisciplinary research, where the same concerns about accuracy and bias may apply. For example, State et al. study international migration among Yahoo! users and its economic determinants; they note that "though IP-based geolocation has been found to be very noisy at the city level, it is generally considered reliable at the country level" (State, Weber, and Zagheni 2013). Unlike most MOOC researchers, they take care to "safeguard against noisy geolocations due to proxy servers" by removing users with unrealistic recorded movement patterns. Zagheni and Weber similarly track migration using e-mail data (Zagheni and Weber 2012). Mielke and Chen use IP geolocation to locate the hosts of Jihadist websites, sometimes to the city level (Mielke and Chen 2008). Hoffman et al. use "hyperlocal" geolocation to improve community event calendars (Hoffman et al. 2012).

More broadly, IP geolocation - frequently on a granular scale - is used in a tremendous variety of applications across many sectors. Institutions use it to identify subscribers (Goodell and Syverson 2007), media companies to display local news (Shavitt and Zilberman 2011), financial companies to block fraud (Curran and Orr 2011), and technology companies to target advertising (Shavitt and Zilberman 2011) or prevent spam and abuse (Goodell and Syverson 2007). Retailers use it to customize prices (Valentino-DeVries, Singer-Vine, and Soltani 2012) or apply e-commerce taxes (Svantesson 2019), despite professional recommendations. Police use it to track down criminals, sometimes with disastrous consequences for innocent people (Solon 2016). Federal agencies may use it to prevent cyber warfare (Shavitt and Zilberman 2011).

## 5.4. IP geolocation technology

IP geolocation databases rely on Internet Protocol addresses, numerical identifiers assigned to devices (RIPE Network Coordination Centre 2019). Throughout this paper, I use addresses from the IPv4 (Version 4) system, an older assignment scheme which is currently being replaced with IPv6. An IP block is a group of consecutive IP addresses - anywhere from a single address to billions of them. The addresses in a block are typically assigned to the same Internet Service Provider (ISP) or other organization, which suggests, but certainly does not prove, that the associated devices are physically near each other.

Correctly identifying a user's location from an IP address is in principle a difficult problem. ISPs can reassign addresses to different organizations or devices frequently, arbitrarily, and confidentially. A mobile devices' address changes as a user moves, and a laptop's IP address can change each time a user connects to a new Wi-Fi network.

Furthermore, even correctly tying an IP address to a device location does not necessarily correctly place the user. Many technologies can obscure locations: for example, someone using AOL dial-up Internet in a rural area, a Secure Shell to connect to work remotely, a virtual proxy network to bypass the Great Firewall of China, or Tor to conceal her activity can present with an IP address that belongs to a device located far away from her (Nitke vs. Ashcroft - Declaration of Seth Finkelstein 2004; Muir and van Oorschot 2006).

Each entry of an IP geolocation database like MaxMind's links an IP block to a physical location such as a country, state, city, ZIP Code, or approximate latitude/longitude. MaxMind does not publicize the sources of its data, nor details about its evaluation methods.

Muir and van Oorschot 2006 break down the techniques that exist for IP geolocation and their limitations, which I summarize below. Presumably, MaxMind populates its database using some combination of these methods. Muir and van Oorschot also discuss the possibility of intentional evasion of geolocation; a user who wants to obscure his location has many effective tools at his disposal.

In some circumstances, the owners of IP blocks publish contact information in public registries, and locations can be inferred from the phone numbers and mailing addresses. These self-submitted records can be incorrect, out of date, or unlocalized - an IP block's owner can be very far from a particular device. Similar registries (and issues) exist for autonomous systems, organizational networks which advertise the IP addresses that they include, and for domain names, like *mit.edu*.

Some other data are easily available but less systematic. Occasionally, administrators choose to include approximate geographic coordinates in responses given by servers that translate domain names to IP addresses. Sometimes information can be extracted from domain names: one can guess the location of the server for *www.city.cleveland.oh.us* (though the guess might well be incorrect, since sites can be hosted on faraway servers).

Sometimes users provide clues to their location, as by entering a shipping address, looking up a local news site, or specifying a language and timezone in browser settings.

An IP address's location can be measured more directly by sending messages to it from different locations and tracking the response times - a similar idea to the trilateration used by GPS and cell phones. This active approach is likely more costly, and it requires the measured device to cooperate. A workaround involves tracing the route that a message takes among different servers. Consecutive servers in the route should be geographically nearby; if some locations are known, one can roughly infer the others.

I suspect that MaxMind primarily uses a final, simplest approach: buying information from ISPs (or tricking them into revealing it). ISPs know their network configurations and policies for assigning addresses; the more ISPs with which a geolocation company maintains a relationship, the more accurate its database.

## 5.5. IP geolocation accuracy

Despite the many uses of IP geolocation in practice, networking literature evaluating database accuracy is consistently unfavorable.

Shavitt and Zilberman 2011 examine several IP geolocation databases by comparing them to groups of IP addresses that are known to share a physical location, noting that "evaluating the accuracy of these mapping services is complex since obtaining diverse large scale ground-truth is very hard." They find that "none of the databases is close to its acclaimed accuracy on the country level." MaxMind in particular disproportionately places IP addresses to the U.S, and especially to Washington, D.C. More generally, although most geolocations are correct, "there is a long and fat tail of errors in the databases; These errors are in the range of thousands of kilometers and countries apart. The use of geolocation databases should therefore be careful."

Poese et al. 2011 similarly find that database entries overrepresent the U.S. Using ground-truth data from a large European ISP, they conclude that "in most of the cases [...], the location given by the databases is off by several hundreds, even thousands of kilometers" and that "geolocation databases can claim country-level accuracy, but certainly not city-level."

Siwpersad, Gueye, and Uhlig 2008 compare database entries with delay-based active measurements. They conclude that by tying an entire IP block to a single location, databases "are too coarse to provide accuracy at the level of individual IP addresses". They also express general skepticism: "Geolocation databases should either provide information about the expected accuracy of the location estimates within each block, or reveal information about how their location estimates have been built, unless databases have to be trusted blindly." More precisely, Gueye, Uhlig, and Fdida 2007 find that half of MaxMind's blocks span devices located more than 500 kilometers apart, which makes it impossible to accurately link the the whole block to one city.

## 5.6. Geographic proxies for socioeconomic status

In public health research, several studies have examined whether local economic measures are an effective proxy for individual socioeconomic status; results are mixed.

Geronimus and Bound 1998 find that ZIP Codes are often-inaccurate and biased markers of individual socioeconomic status: for example, "there is 89 percent as much variation in income within zip codes as in the general population". They warn "that aggregate measures can not be interpreted as if they were microlevel variables".

Similarly, Soobader et al. 2001 find that ZIP Codes and census blocks are both downwardly biased measures of individual socioeconomic status. ZIP Codes are less biased than census blocks in measuring individual income, but more for education. They conclude that "Researchers should be cautious about use of proxy measurement of individual SES even if proxies are calculated from small geographic units."

Conversely, studying Canadian neighborhoods Mustard et al. 1999 find evidence in support of using neighborhood-level "measures of income in studies which do not have access to individual-level income measures."

# 6. Summary of datasets

My work relies on combining several sources of data, which I describe in the sections below and illustrate in Figure 1. My data on users comprise edX registration records (6.1), which include users' IP addresses, and mailing addresses from surveys provided by some users (6.2). I use the MaxMind IP geolocation database (6.3) to translate users' IP addresses to ZIP Codes and the Google Maps Geocoding API (6.4) to translate mailing addresses to ZIP Codes. My ZIP Code data comprise the DCI economic dataset (6.5), which includes a distress score from 0 to 100, and the Census Bureau shapefiles (6.6), which include geographic data. In 6.7, I comment on limitations of the datasets.

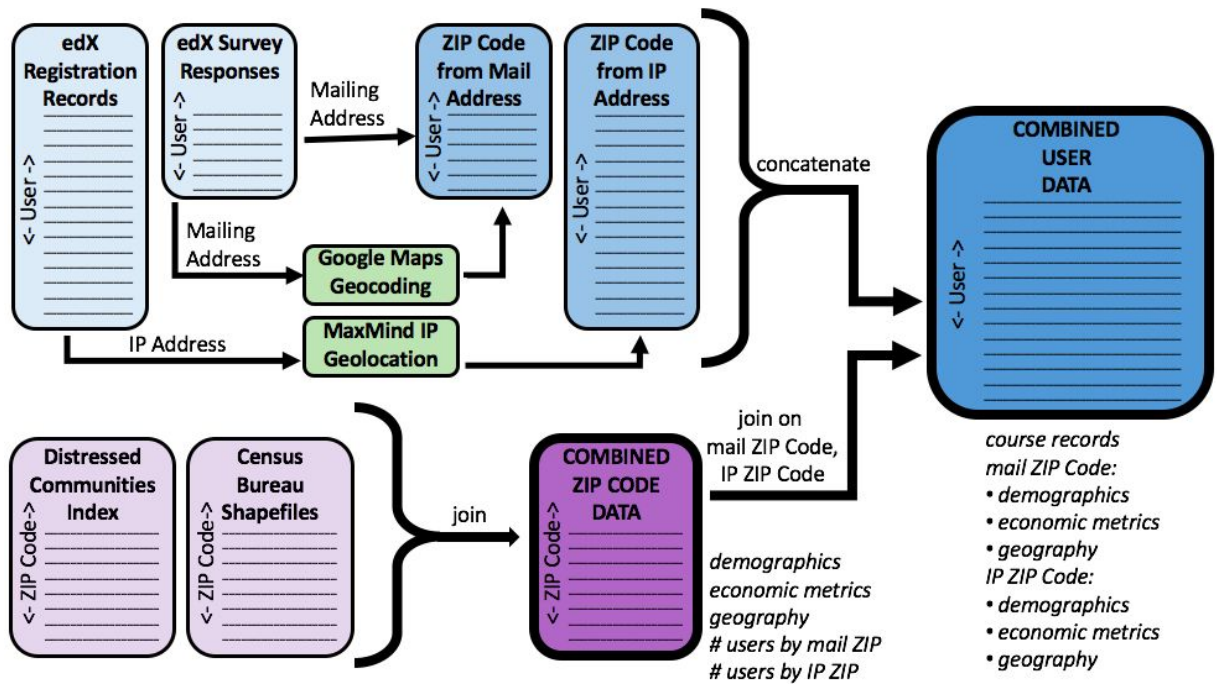

**Figure 1: Combining datasets.** Shows how I combined datasets; sections 7.2 and 8.2 provide more detail. I concatenated data on users (edX registration records and survey responses), used Google Maps Geocoding to obtain a ZIP Code from the mailing address, and used MaxMind to obtain a ZIP Code from the IP address. For ZIP Codes, I combined demographic and economic data from DCI with geographic data from the Census Bureau shapefiles. For each user, I joined in data about both ZIP Codes (from mailing address and from IP address).

## 6.1. edX registration records

The MOOC dataset (Lopez 2016) contains information on all U.S.-identified registrations for HarvardX and MITx courses from 2012 through the early months of 2018. A *registration* refers to a user-course pair; a single user cannot have multiple registrations for the same course, but can have many registrations across different courses.

The record for each registration contains, among other data:

- Information on the course: title, start and end date, hosting institution (Harvard or MIT), subject category, minimum passing score.
- Automatically recorded information on the user: user ID, email address domain, some browser settings, whether the user is course staff.
- User demographic data, if the user reports them: year of birth, gender, level of education.
- Modal IP address: the most common IPv4 address from which this user has connected.
- Summary information on the user's engagement with the course: whether the user has viewed any course materials, completed the course, verified an identity, and earned a certificate; the user's final numerical grade in the course.
- Many granular records of the user's engagement with the course: for example, how many times the user has played and paused videos, commented on the course forums, or clicked "Show Answer" for problems; the average time between interactions with the system.

In all analyses, I use only those registrations where the user is not identified as course staff.

## 6.2. Mailing addresses from edX Survey Responses

When registering for MITx or HarvardX, users are asked to fill out a survey that includes a free-form text box asking for a mailing address. Of the users that edX identified as being in the U.S., 135,000 users provided a response. Many of the addresses are incomplete (for example, including a street address but no city or vice versa), outside the U.S., or clear mistakes, such as email addresses or text responses to a different survey question.

## 6.3. MaxMind IP geolocation dataset

MaxMind is a commercial provider of IP geolocation datasets. I use the September 19, 2018 version of its mid-tier GeoIP2-City dataset, which maps IPv4 addresses to postal codes (MaxMind, Inc. 2018).

MaxMind publishes statistics on the accuracy of its dataset, "calculated by checking known web user IP address and location pairs against the data" (MaxMind, Inc. n.d.). The provenance and distribution of the "known" data are not specified. As of May 2019, MaxMind claims that the GeoIP2-City dataset maps an IP addresses to the correct ZIP Code 36% of the time, to a location within 10 km of the correct one 62% of the time, and to a location within 100 km of the correct one 90% of the time.

## 6.4. Google Maps Geocoding API

The Google Maps Geocoding API (Google Developers 2019) takes as input an address, which may be imperfectly formatted, and returns a well-formatted address, ZIP Code, latitude/longitude, and information on the estimated accuracy and precision of the result. The confidence metrics are not optimally helpful, for several reasons:

- The API always returns at most one result, even when there are many possibilities. For example, in the U.S. there are many places located at "123 Main St." but just one located at "9710 Chicken Dinner Rd" (Chatter 2008). The API returns a single address for each, without any indication that there could be many equally valid responses in the first case.
- The result may include a "Partial" indicator when the returned address only partially matches the input address, but I have observed cases without the indicator in which the two addresses are clearly different.
- When the input address includes an apartment number, the API may erroneously parse it as a street number.

Fortunately, in the vast majority of cases I observe, the output address and input address seemed to correspond. I mitigate the first concern by only inputting addresses to the API that appear to include either a ZIP Code or both a city and state, which suggests uniqueness.

## 6.5. Distressed Communities Index (DCI) dataset

The Economic Innovation Group (EIG) produces a dataset measuring the economic health of more than 25,000 U.S. ZIP Codes (Economic Innovation Group 2018). I use the 2012-2016 version. For each ZIP Code, the dataset includes its population; its proportion of residents of various ethnicities; its population density (on a scale of "Low" to "Very High", which I code as 1 to 4); and seven economic metrics, whose definitions are simplified here: its proportion of adults without a high school diploma, housing vacancy rate, unemployment rate, poverty rate, ratio of median income to state median income, percent change in employment from 2012 to 2016, and percent change in number of businesses from 2012 to 2016.

Most importantly, the dataset includes a composite Distressed Communities Index (DCI) for each ZIP Code. To calculate the DCI, the EIG ranks the ZIP Codes for each of the seven metrics, then averages and normalizes the ranks to reach a final composite score. DCI scores are distributed uniformly from 0.0 (best) to 100.0 (worst) over all ZIP Codes, and each ZIP Code is placed in a corresponding tier: "prosperous" (DCI of 0-20), "comfortable" (20-40), "mid-tier" (40-60), "at risk" (60-80), and "distressed" (80-100).

## 6.6. Census Bureau ZIP Code shapefiles

The Census Bureau distributes shapefiles (U.S. Census Bureau 2012a) for ZIP Code Tabulation Areas, which are polygons that approximate ZIP Codes (see 6.7). I use the 2018 version, which is based on the 2010 census. For each ZIP Code, the shapefiles include its land and water area (which I combine to use as total area), a list of latitude/longitude coordinates laying out the vertices of the polygon, and the coordinates of the internal point (U.S. Census Bureau 2012b). The internal point is the same as the geographic center in most cases, except when the geographic center lies outside the polygon - one could imagine a crescent-shaped ZIP Code. In that case, the internal point is the closest point to the geographic center within the boundaries of the polygon (U.S. Census Bureau 2018).

## 6.7. Dataset limitations

ZIP Codes are more complicated than they may first appear (Hurvitz 2008). A ZIP Code is not actually a geographic area, but a proprietary set of mailing addresses or Post Office Boxes determined by the U.S. Postal Service. A ZIP Code Tabulation Area (ZCTA) is a polygon that approximates the area that contains those addresses, as determined by the U.S. Census Bureau. Unfortunately, the two do not always correspond, and ZIP Codes can change frequently, making ZCTAs go out of date. The different datasets I use variously use ZIP Codes or ZCTAs (DCI uses both), and the documentation does not always specify which. I use "ZIP Code" to refer to either throughout this paper, but the distinction certainly introduces some error into my calculations.

I also note that the datasets used come from different years, and it is likely that some addresses, IP address assignments, ZIP definitions, and economic/demographic profiles shifted over time.

# 7. Psychometric analysis of course engagement rates by ZIP Code

In this chapter, I describe my work studying economic patterns in edX usage according to IP-geolocated ZIP Codes. I overview my objectives (7.1) and approach (7.2). I then explain my results (7.3), including a general overview of the data (7.3.1), an analysis of economic relationships (7.3.2), an analysis of per-course registration patterns using psychometric methods (7.3.3), and a similar analysis for certification and completion rates (7.3.4). I discuss the implications and limitations of this analysis in 7.4.

## 7.1. Objectives

In this section, I study engagement with HarvardX and MITx courses in different U.S. ZIP Codes, identifying ZIP Codes with IP geolocation.

To begin, I investigate the relationships between per-capita usage rates and ZIP Code economic factors. I aim to replicate the findings of (Hansen and Reich 2015b) using a more familiar geographic unit, a more comprehensive measure of economic prosperity, and a larger sample of users - nearly all U.S.-identified edX users for six years, rather than a (possibly unrepresentative) set of survey respondents. A consistent finding would also lend support to the use in future research of fine-grained IP geolocation, which is more easily scalable than Hansen and Reich's painstaking mailing address parsing.

I also work toward a ZIP Code "MOOC propensity" score: a measure of inherent orientation towards MOOC enrollment. Compared with a simple metric like total registration count, such a score might evince stronger relationships with economic factors and be less sensitive to random noise or unusual events, such as high enrollment in a particular course in a particular ZIP Code because of a local university assignment. It could be useful to researchers studying economic trends in MOOC participation, or MOOC providers who wish to target marketing to lower-propensity areas or identify courses with atypical registration distributions.

In searching for a score of MOOC propensity, I perform exploratory factor analysis using ZIP Codes as subjects and per-course registrations as observable variables. Using geographic areas as subjects in factor analysis introduces some unique issues, including potential dependence between adjacent areas; to mitigate this concern, I note the unidimensional nature of the data.

I operationalize MOOC engagement using the per-capita number of registrations, course completions, and certifications in each ZIP Code. The latter two constructs are similar: completing is a prerequisite to certifying, but users who complete courses may need to verify their identities to receive certificates; these rules have differed across courses and time. Many other metrics are possible to capture a more detailed notion of participation, such as using forum posts or content views (DeBoer et al. 2014) or asking for user self-reports (Hone and El Said 2016). Another interesting research avenue involves studying certification and completion rates among those who have registered, rather than among all ZIP Code residents. This question is less suitable to address with factor analysis because of extensive missing data - it's difficult to define a completion rate if there are no registrants in a particular ZIP Code for a particular course, as is very often the case.

I aim to study the geographic distribution of MOOC engagement and the associated economic factors, addressing the following questions:

- What is the distribution of rates of course registration, completion, and certification over U.S. ZIP Codes?
- What are the relationships between ZIP Code course engagement rates and ZIP Code economic properties?
- Are ZIP Code course engagement levels unidimensional? If so, could this trait constitute a "MOOC Propensity" which is more helpful than simple registration counts in predicting economic factors?

## 7.2. Approach

A key preliminary step for all analyses is combining information from the three different datasets. For visualizing distributions of course engagement, I make maps using Python's Pandas, GeoPandas, Matplotlib, and PySAL libraries. I perform quantitative analyses and generate figures using Python and Stata.

In the bulk of this work, I apply factor analysis to study course engagement at the level of ZIP Codes. I model MOOC engagement as an exam in which the test-takers are ZIP Codes, the items are individual courses, and a ZIP Code's score for a particular course is the number of registrations (or completions, or certifications) per million ZIP Code residents.

In particular, restating the explanation from (The Pennsylvania State University n.d.), I have $p$ courses. Each data point $r_{jk}$ is the number of registrants for course $j$ in ZIP Code $k$ per million ZIP Code residents. I let $u_j$ be the mean across ZIP Codes for course $j$. I aim to fit $m$ factors, finding a factor loading $l_{ij}$ for each factor $i$ for each course $j$ and a factor score $s_{ik}$ for each factor $i$ for each ZIP Code $k$, so that each $r_{jk} = u_j + \sum_{i=1}^{m} l_{ij} s_{ik} + e_{jk}$, where $e_{jk}$ is an error term that is to be minimized.

I use classical test theory, including Cronbach's alpha and item-level statistics, to assess item reliability. After performing factor analysis, I use Scree plots to assess unidimensionality. I look for correlations between factor loadings, factor scores, and other variables after applying transformations to attempt to linearize relationships. In particular, for ZIP Codes, I look for relationships among first-factor scores, number of completions and certifications per population and per number of registrants, DCI, the seven component factors of DCI, percentage of minority residents, population, and population density. For courses, I look for relationships between factor loadings, number of registrations, completions, and certifications, and percent of registrants who completed or certified. I typically note correlations with $|r| > 0.2$; I do not assess the statistical significance of relationships.

## 7.3. Results

### 7.3.1. Data summary

In the psychometric analyses, I select only those registrations where MaxMind geolocates the IP address to a U.S. ZIP Code for which Distressed Communities Index data are available. This leaves ~2.7 million registrations distributed over ~18,000 ZIP Codes and 576 courses, many of which are iterations of the same course in different semesters.

Figure 2 and Figure 3 illustrate the distribution of MOOC registrations over ZIP Codes, as identified by MaxMind geolocation.

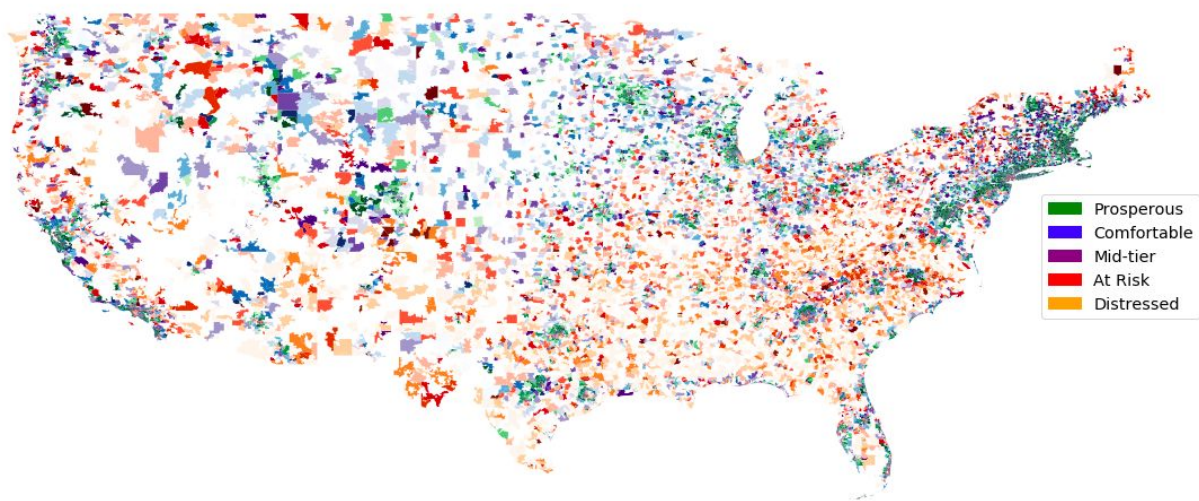

**Figure 2: Registration map.** HarvardX/MITx registrations from 2012 to 2018 by DCI tier. Darker colors indicate a higher number of registrants per capita.

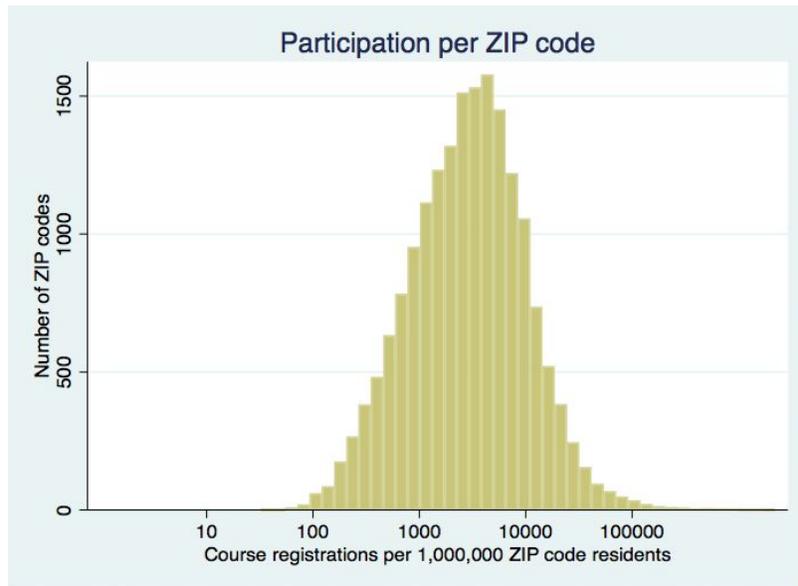

**Figure 3: Histogram of ZIP Code registration rates.** Distribution of per-capita HarvardX/MITx registration rates over ZIP Codes.

ZIP Codes in this dataset, which excludes ZIP Codes with no identified registrations, had between 32 and 2.2 million registrations per million population, with a median of 2,970 and a mean of 7,380. A single user may have multiple registrations for different courses, so the numbers above 1 million are suspicious (remembering that geolocation has limited accuracy) but not impossible. The distribution is heavily skewed toward low counts; note the logarithmic scale in Figure 3.

It is interesting to consider the ZIP Codes with above 900,000 registrants per million population: they lie in Cambridge, MA; Salt Lake City, UT; Lincoln, MA; Manhattan, New York City, NY;

St. Louis, MO; Seattle, WA; and Boston, MA. Most of these areas are well-known tech hubs, and three are near the Cambridge home of Harvard and MIT.

Some ZIP Codes have small recorded populations, which makes them sensitive to inflated registration rates. Of the ~18,000 ZIP Codes, populations according to DCI data range from 510 to ~115,000, with a median of 10,860 and a mean of 16,535.

The distribution of course registrations has a similar skewed shape (Figure 4). The smallest course has just 12 registrations, while the largest has ~164,000 (median 2,067, mean 4,649). The largest courses are several iterations of introductory computer science courses from both MIT and Harvard.

In total, out of 10.5 million possible combinations of a course and a ZIP Code, 1.1 million are seen in the data. Figure 5 shows the distribution of registrations among these 1.1 million pairs. Within this group, registration rates range from 8.7 to ~220,000 registrations for a single course per million ZIP Code residents, with a median of 52 and a mean of 121.

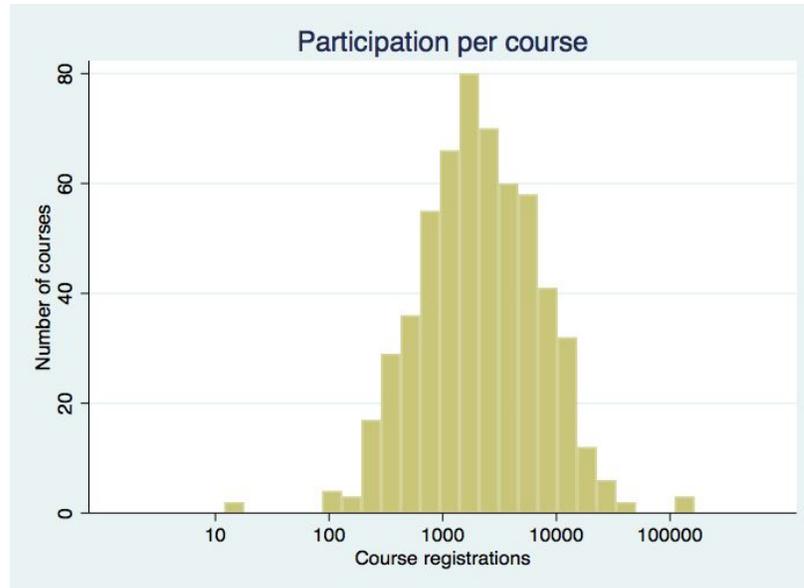

**Figure 4: Histogram of course registrations.** Distribution of number of registrations by course; note log scale.

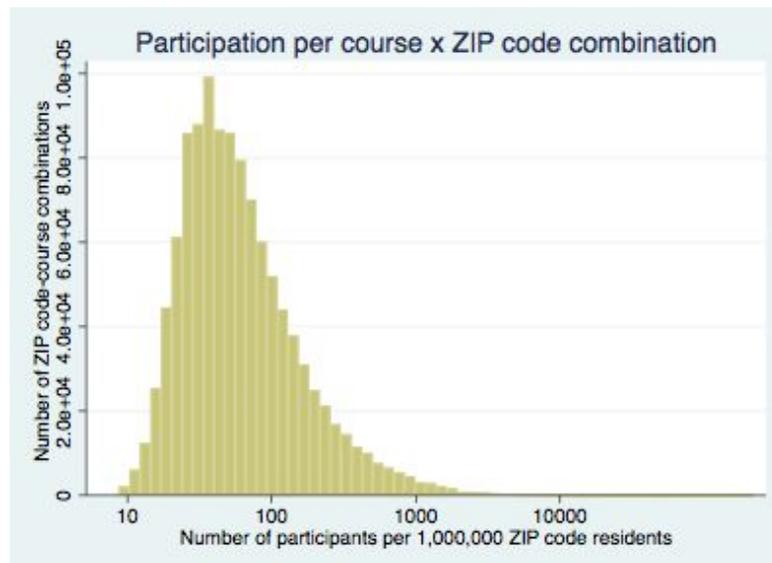

**Figure 5: Histogram of course-by-ZIP registrations.** Distribution of registrations for course-ZIP Code pairs among the 1,110,765 observed; note log scale.

## 7.3.2. Economic relationships

As expected given previous literature, more prosperous ZIP Codes have more registrations per capita (Figure 6). Across ZIP Codes, the logarithm of per-capita registrations correlates positively with population density (r = .36) and median income ratio (r = .24) and negatively with proportion of adults without high school diplomas (r = -.27), DCI (r = -.26), and unemployment rate (r = -.25).

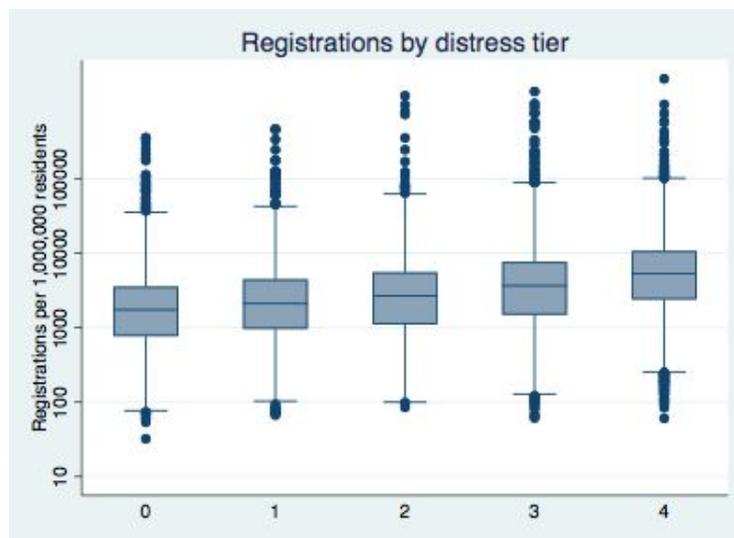

**Figure 6: Boxplot of ZIP Code per-capita registrations**. Tiers from 0 (distressed) to 4 (prosperous). For each tier, the box depicts the 25th, 50th, and 75th percentile, the brackets depict adjacent values (those within a fraction of the interquartile range on a log scale), and the dots represent outlying values.

The relationship is less clear between prosperity and the proportion of registrations that result in completion or certification, perhaps because across the board fewer than 4% of registrations end in completion. The shifted logarithm of completions per population correlates negatively with logarithm of population (r = -.20), as does the shifted logarithm of certifications per population (r = -.21). The shifted logarithm of per-registration completion rate correlates negatively with logarithm of population (-.53), population density (-.30), and proportion of minority residents (r = -.23). Results for per-registration certification rate are very similar (the same correlations are -.53, -.30, and -.22, respectively). Interestingly, these results imply that people from more dense areas are more likely to enroll in courses, but that registrants from more dense areas are less likely to complete courses. These results may suggest network effects in which more connected people have greater exposure to MOOCs, but no more motivation to complete them.

Discussing these patterns is difficult because of the shapes of the distributions; reducing a ZIP Code's per-registration completion rate to a single number is nontrivial. The distribution of per-registration completion rates is log-normal, except for a huge spike at zero (Figure 7): nearly half of ZIP Codes have no completions. This suggests that there are two different processes determining a ZIP Code's completion rate; understanding these processes is a question for future research. If we consider the means of raw per-registration completion rates in a tier, or their shifted logarithm, more prosperous tiers come out ahead (Figure 8). If we consider an unshifted logarithm (excluding ZIP Codes with zero completions), more distressed tiers come out ahead. I suspect that the latter effect is simply because more distressed ZIP Codes have fewer registrations, so the denominator in the ratio is smaller.

Appendix 11.1 includes all the correlations discussed in this section.

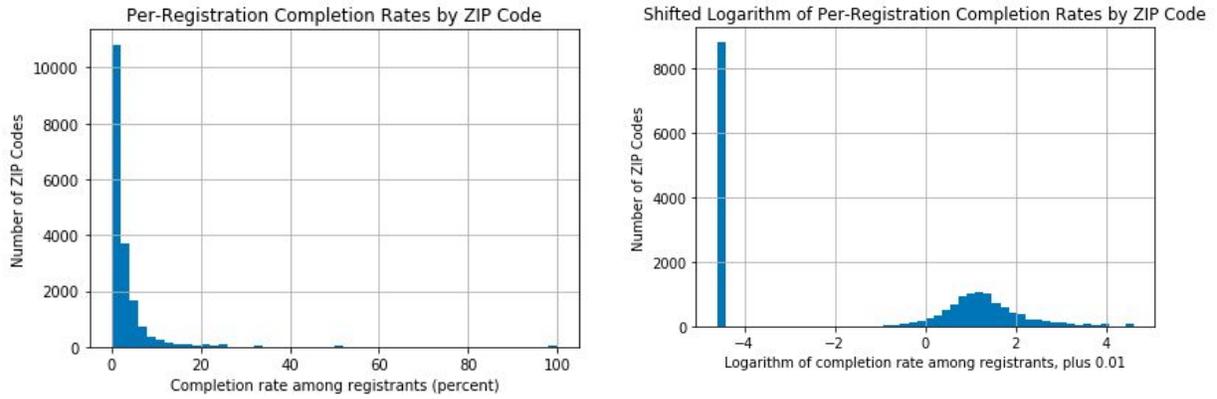

**Figure 7: Histograms of per-registration completion rates.** The left graph shows the distribution of per-registration completion rates, and the right graph the logarithm of these rates plus 0.01. The large spike at the left of each graph shows the ZIP Codes with no completions.

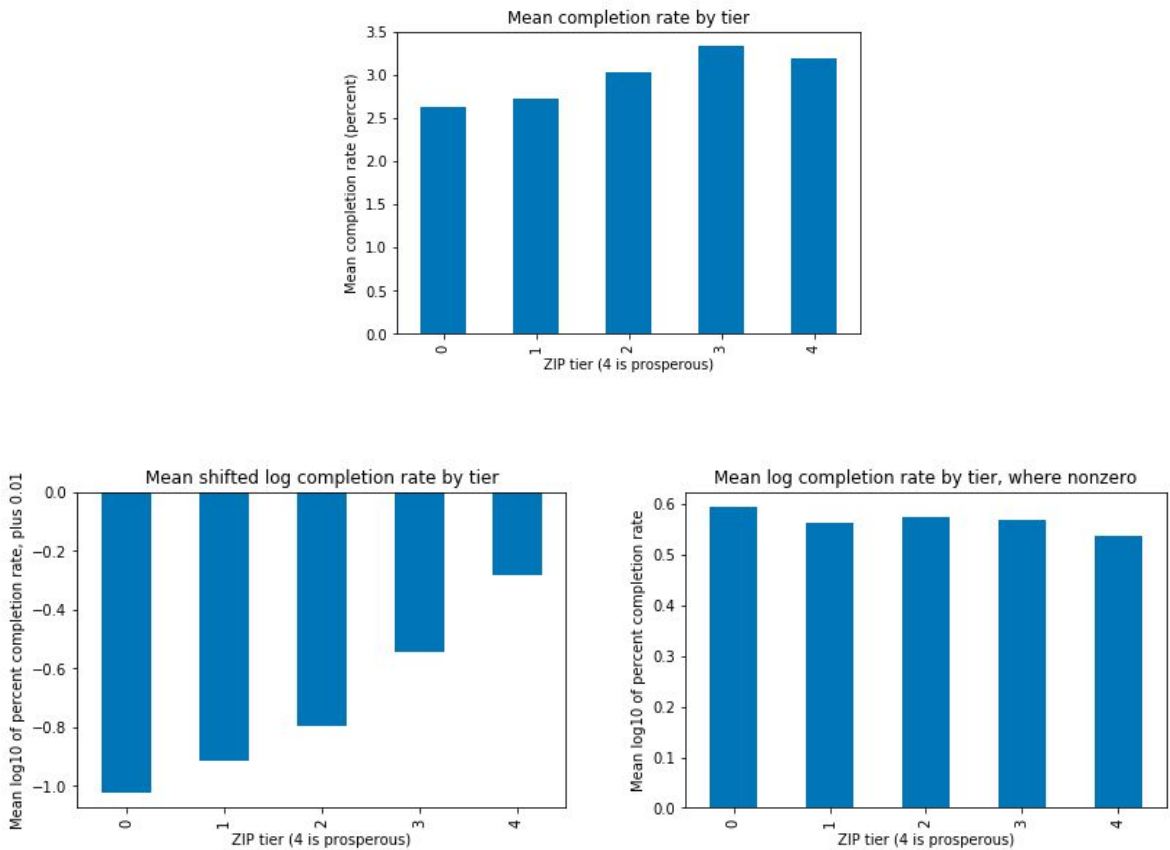

**Figure 8: Mean ZIP Code completion rates across tiers.** Top, raw completion rate; left, shifted log; right, mean log, where nonzero.

### 7.3.3. Psychometric analysis with registrations

I apply factor analysis to look for a ZIP-level "MOOC propensity" score. By taking advantage of the information captured in the distribution of a ZIP Code's registrations over courses, such a score could potentially track more closely with economic factors.

Treating ZIP Codes as persons, courses as items, and per-million registrations as scores, I find a high Cronbach's α (UCLA Institute for Digital Research & Education n.d.) reliability of 0.98. To aid the interpretation of this number, the Spearman-Brown formula prophecies a reliability of 0.64 if the number of courses were reduced to 20. Factor analysis suggests unidimensionality; a single factor explains a large plurality of variance (Figure 9). I experiment with different operationalizations of the data, such as dichotomizing or taking the shifted logarithm, and find similar effects: in both cases, Cronbach's α of 0.99, prophesied to .86 for 20 courses, and distinct unidimensionality visible in the Scree plots. I use untransformed registrations per million population in further analysis.

I also examine item statistics for each course. Item-test correlations vary from .05 to .89, and are strongly correlated in turn with course registrations (r = .80) (Figure 10).

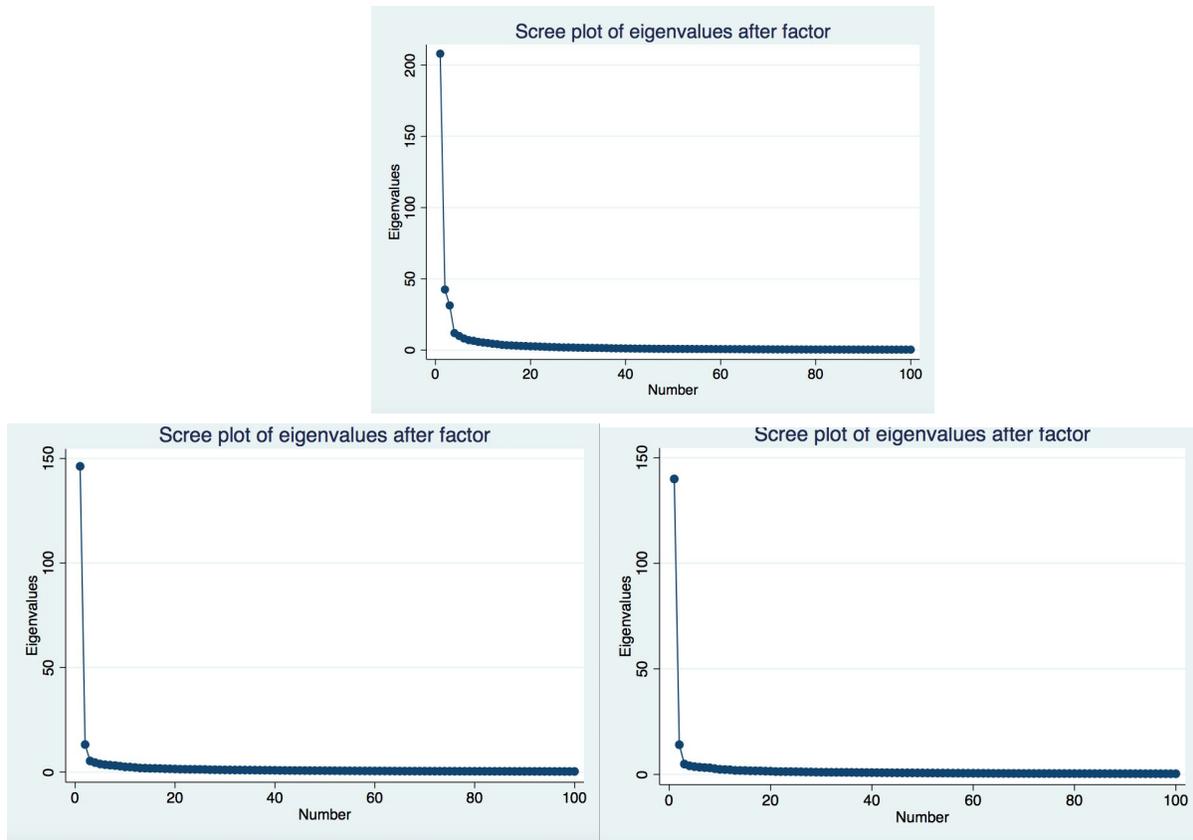

**Figure 9: Scree plots.** Plots of eigenvalues after performing principal-factors factor analysis. The single outlying point suggests that a single factor dominates. The data are course registrations per million - at top, untransformed; at left, dichotomized to 0 (no registrations) or 1 (any registration); at right, transformed by adding 1 and taking the logarithm.

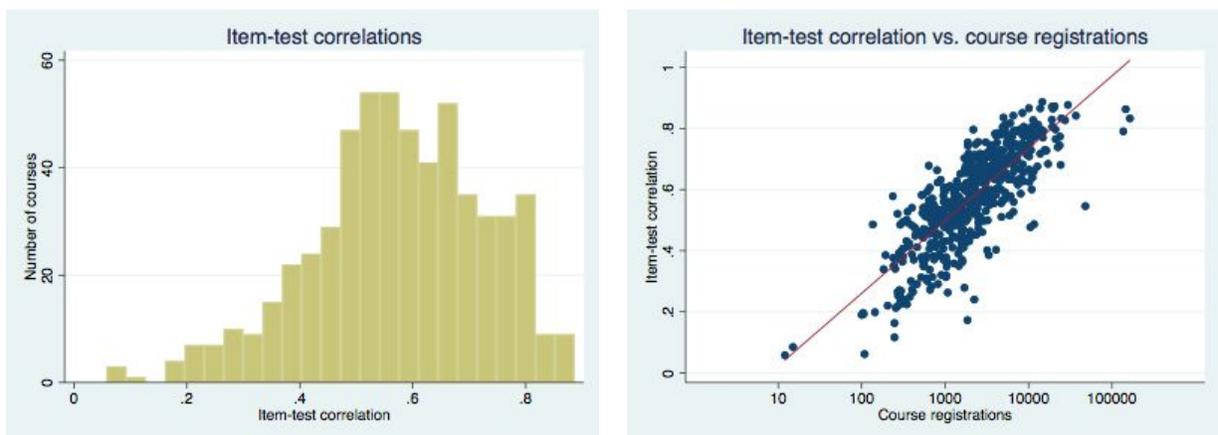

**Figure 10: Item-test correlations.** Left, histogram of item-test correlations; right, scatterplot of item-test correlations against course registrations.

The courses' first-factor loadings correlate strongly with the logarithm of course registration counts (r = .72) (Figure 11), and near-perfectly with the item-test correlations (.98) - which makes sense given that the data are unidimensional; the more registrants a course has, the more information it gives about a ZIP Code. Similarly, item uniqueness (the percentage of variance in a course's registration left unexplained by the first factor; Figure 12) is highly negatively correlated with the factor loadings (r = -.98) and with item-test correlation (r = -.97). Overall, the first factor explains 46% of all variance.

All the correlations discussed above are available in Appendix 11.2.

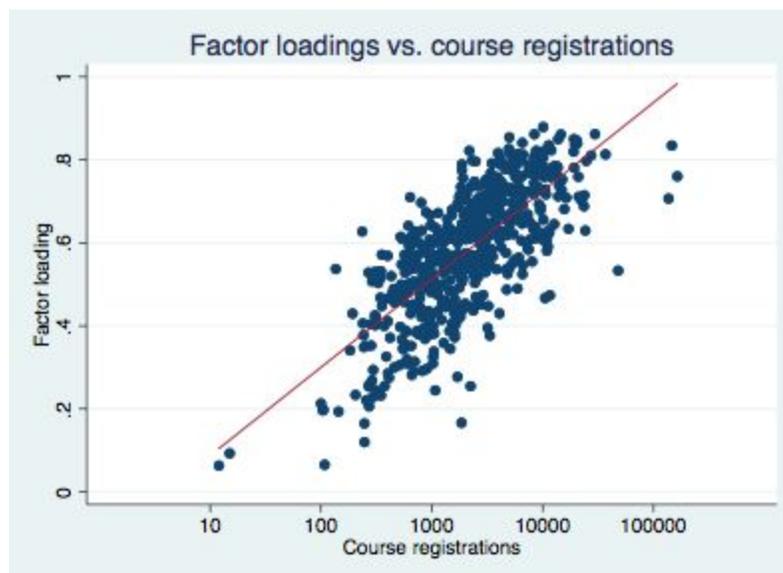

**Figure 11: Scatterplot of factor loadings vs. logarithm of course registrations.**

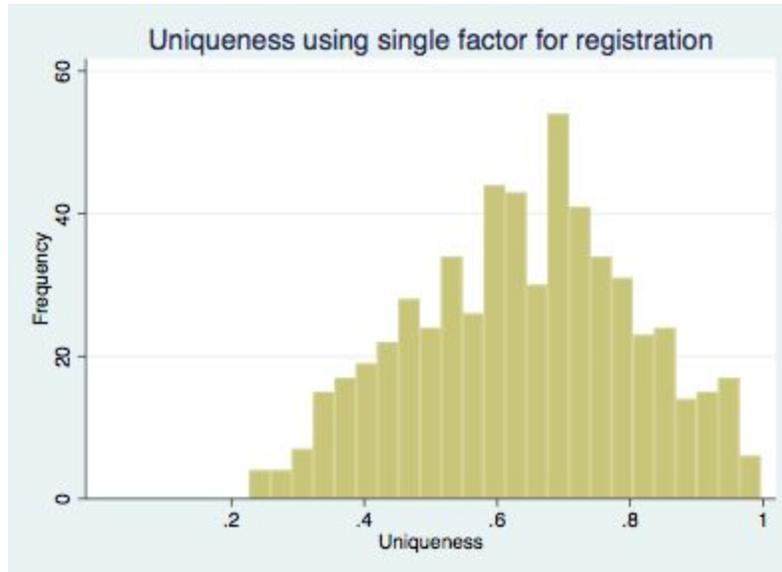

**Figure 12: Histogram of course uniqueness (with respect to the first factor).**

Moving our attention to ZIP Codes, a transformed version of first-factor scores correlates very strongly with the logarithm of per-capita ZIP Code registration (r = .99; Figure 13). Furthermore, correlations between first-factor scores and economic variables are very similar to the correlations using per-capita registration discussed above. Together with unidimensionality, these results suggest that factor analysis does not produce better insight than treating ZIP Code registrations and course registrations as independent. Roughly speaking, Stata estimates a particular ZIP Code's enrollment in a particular course by estimating that ZIP Code's total registrations across all courses and then scaling it according to that course's estimated proportion of all registrations.

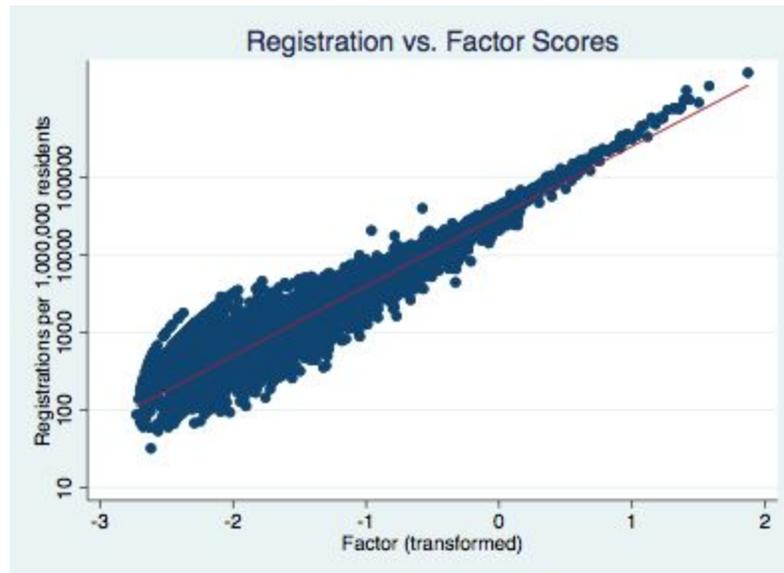

**Figure 13: Scatterplot of registration vs. factor scores.** Logarithm of per-million registrations against the shifted logarithm of factor scores.

### 7.3.4. Psychometric analysis with completions and certifications

I repeat the above analyses using both number of completions and number of certifications per million residents rather than number of registrations per million residents. The results of these analyses are also included in Appendices 11.1 and 11.2; factor scores and loadings are labelled as coming from "registration", "completion", or "certification" accordingly.

Factor analysis drops some courses that have zero variance (no completions). Among the remaining courses, Scree plots indicate a reasonable degree of unidimensionality, particularly for transformed data (Figure 14). The first factor has less explanatory power here; uniqueness for most courses is near 1 (Figure 15).

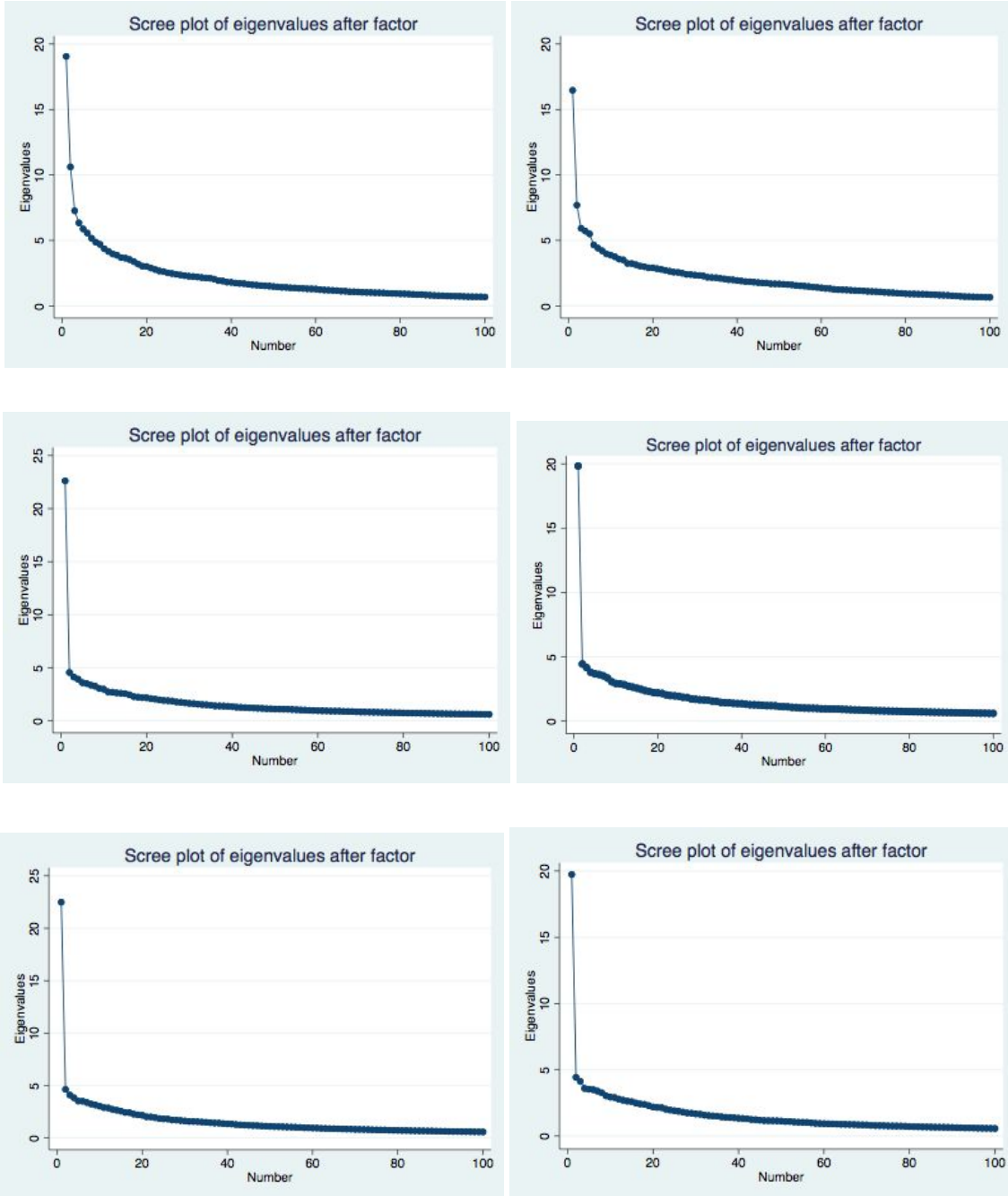

**Figure 14: Scree plots for completions (left) and certifications (right).** Raw data (top), dichotomized data (middle), and shifted logarithm of data (bottom).

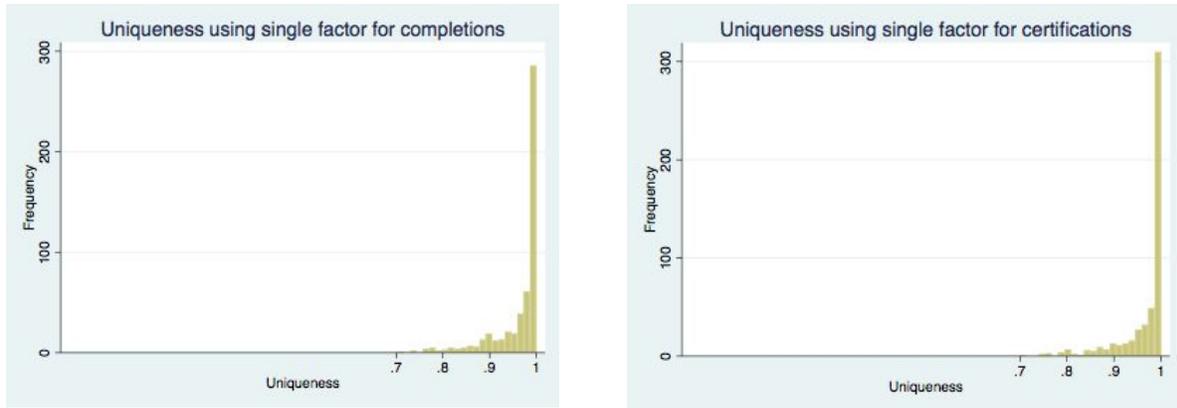

**Figure 15: Histograms of uniqueness.** Distribution of uniqueness of courses for completions (left) and certifications (right).

Logarithmically transformed first-factor scores for completions correlate highly with log of number of completions per million (r = 0.87), and analogously for certifications (r = .86). The transformed first-factor scores correlate similarly with economic variables, as do the corresponding raw data. Although the Scree plots indicate the possibility of a meaningful second factor, the second factor does not correlate with any economic variables.

## 7.4. Discussion

In this chapter, I use factor analysis to calculate a MOOC propensity capturing ZIP Codes' orientation towards MOOCs. I find that whether calculated using registrations, completions, or certifications, such a propensity captures little more than the raw per-population rate in terms of relationships with economic variables. Although it does not offer new socioeconomic insights, this propensity could still be useful for studying new ZIP Codes or courses about which little information is available.

The high reliability and degree of unidimensionality of the data, as well as the correlation of both factor loadings and scores with enrollment counts, tell a consistent story that individual courses behave much alike in terms of their geographic distribution. This is somewhat surprising: one might expect that some courses would appeal more to people in some areas and economic situations than others. It is possible that these trends do exist, but do not manifest in this analysis because of the large number of ZIP Codes compared to registrants - most courses have no enrollments in most ZIP Codes. They might emerge if I repeated this analysis after grouping courses together by time period or topic, or grouping ZIP Codes together by geographic proximity or economic similarity. When grouping by subject category, I saw some preliminary evidence that STEM and humanities courses are slightly more popular in more distressed areas, and computer science courses in more prosperous ones.

These analyses rely on ZIP Codes derived from IP geolocation, which may be substantially inaccurate. In the next chapter, I study the accuracy of these geolocations.

# 8. Accuracy and bias of IP geolocation

In this chapter, I describe my study of IP geolocation accuracy and bias among edX users. In 8.1, I explain my objectives. In 8.2, I explain my approach to extracting ZIP Codes from mailing addresses (8.2.1), finding ZIP Codes from IP addresses (8.2.2), calculating geolocation error (8.2.3), and performing analyses (8.2.4). In 8.3, I explain my results, including general observations (8.3.1), economic patterns in registrations (8.3.2), distributions of geolocation error (8.3.3), properties of geolocated ZIP Codes (8.3.4), and overall economic distributions according to geolocation (8.3.5). In 8.4, I discuss implications and limitations of these results.

## 8.1. Objectives

In Chapter 7 I find strong relationships between ZIP-level economic factors and course registrations.

However, this analysis relies on fine-grained IP geolocation - which the networking literature and even MaxMind documentation suggest is, at best, a noisy signal with low precision. Because I look for overall trends rather than accurate identification of individual users, limited precision alone would not post a great problem; it might obscure real trends but not create spurious ones. As a caricature, we could imagine MaxMind assigning each user to a random location within a 20-mile radius of his true position: we would expect the errors to cancel out on average, leaving about the right number of users in each ZIP Code.

However, there remains a more alarming possibility: that geolocation accuracy could be systematically biased by economic factors. It seems plausible that geolocation would make greater errors in poor, rural areas than wealthy, urban ones, due to differences in infrastructure quality and in financial incentives to maintain accurate listings. There is also a possibility of systematically asymmetric misidentifications. Prior research shows that MaxMind preferentially misidentifies users as being in the U.S. and in the Washington, D.C. area (Shavitt and Zilberman 2011); perhaps it also preferentially misidentifies users as being in dense and wealthy areas. A strong such bias could artificially produce the results seen in the previous section, even if true usage were equal across economic tiers.

Finding such biases could threaten the validity of my results from Chapter 7 as well as other research that uses fine-grained IP geolocation, like Alcorn, Christensen, and Kapur 2015. It would also introduce the possibility that commercial and governmental use of IP geolocation inadvertently discriminates against certain user groups. Conversely, demonstrating that IP geolocation performs with low bias and reasonable accuracy would open the doors for fine-grained, inexpensive, easily scalable research on geographic patterns in usage of MOOCs and other Internet services.

To investigate bias, I take advantage of a sample of edX registrants that choose to provide mailing addresses in surveys. I treat ZIP Codes derived from these mailing addresses as "ground-truth" data, and compare them with MaxMind-identified ZIP Codes for the same users. I aim to address several questions:

- To verify the key result of 7.3.2: according to ground-truth data, how do per-capita registration rates vary with ZIP Code economic variables?
- How do the distributions of geolocation error vary with ground-truth ZIP Code economic variables?
- How do the economic profiles of MaxMind-identified ZIP-code vary with the economic variables of ground-truth ZIP Codes?
- How do the overall distributions of users' economic backgrounds differ between ground-truth and MaxMind data?

## 8.2. Approach

### 8.2.1. Identifying ground-truth ZIP Codes

I clean survey responses to remove entries that are clearly not mailing addresses or non-U.S. addresses, leaving ~111,000 entries. Using the usaddress Python library (Deng and Gregg 2014), I parse the mailing addresses to identify the address type, and the city, state, and ZIP Code where included.

I use the Google Maps Geocoding API to obtain addresses' ZIP Code and latitude/longitude coordinates. I geocode only those mailing addresses that include either a ZIP Code or both a city and a state, which I take to indicate a unique location. I use Google geocoding results to clean data and identify a "ground-truth" ZIP Code for each address where available, according to the following rules:

- I consider Google results "confident" when the reported accuracy is at "rooftop" or "range interpolated" levels, rather than "geometric center" or "approximate" (Google Developers 2019). I only ever use coordinates if Google is confident.

- I eliminate addresses that Google confidently places outside the U.S., or for which neither parsing nor geocoding yields a ZIP Code (29% of cases).

- When the Google-reported ZIP Code agrees with the parsed one (63% of cases), I use the ZIP Code and coordinates.

- When parsing yields a ZIP Code but geocoding did not (0.6% of cases), I use the parsed ZIP Code as ground-truth and ignore any coordinates. These addresses are frequently P.O. Boxes.

- When geocoding yields a ZIP Code but parsing does not (4% of cases), I use this ZIP Code as ground-truth. If the usaddress type is "street address", rather than city or P.O. Box., I use the coordinates.

- When parsing and geocoded yield different ZIP Codes (3% of cases), I use the parsed ZIP Code as ground-truth and use the coordinates. These cases often involve well-formed addresses and similar, neighboring ZIP Codes - the differences could be explained by typos or changes in ZIP Code boundaries.

In total, I define a ground-truth ZIP Code for ~79,000 users.

### 8.2.2. Identifying MaxMind ZIP Codes

I identify each user's chronologically first course registration (according to the course's start date) and the corresponding modal IP address. I remove a handful of users who are identified as staff or missing an IP address. I use MaxMind to geolocate the IP addresses, keeping users assigned to U.S. ZIP Codes.

In all my analyses, I focus on the ~76,000 users for whom I identify both a ground-truth ZIP Code and a MaxMind ZIP Code. I use latitude/longitude coordinates for about 73,000 of those.

### 8.2.3. Calculating geolocation error

I joinethe user data with the Census Bureau shapefiles and the DCI economic and demographic dataset, neither of which includes all ZIP Codes.

For the ~69,000 users where data are available, I use the *Shapely* library (Gillies 2018) to calculate a geolocation error I call "boundary distance": the Euclidean distance between the user's coordinates and the nearest point in the MaxMind-identified ZIP Code. If the MaxMind and ground-truth ZIP Codes are identical, this error is 0. Although this error is calculated in degrees latitude/longitude, for interpretability I show results in approximate miles; in the U.S., one degree is (very) roughly 50 miles (Rosenberg 2018).

For a sensitivity check, I also calculate an alternative geolocation error, "centroid distance", for ~71,000 users: the great-circle distance, in miles, between the internal point of the ground-truth ZIP Code and the internal point of the MaxMind ZIP Code.

8.2.4. Analyses

16,000 ZIP Codes appear in ground-truth or MaxMind identifications. For each, I calculate population density by dividing population by total area. I group the ZIP Codes into deciles of population, area, and population density, where data are available (about 14,000 in each case). In some analyses, I group adjacent deciles together into quintiles.

For DCI, I instead use the pre-existing DCI tiers (DCI of 0 to 20, 20 to 40, ..., 80 to 100), and define ten sub-tiers analogously (0 to 10, 10 to 20, ..., 90 to 100). Although the tiers are quintiles of the overall DCI dataset, the ZIP Codes in my data lean towards lower DCI: 28% of ZIP Codes with identified tiers are "Prosperous", 22% "Comfortable", 18% "Mid-tier", 17% "At risk", and 15% "Distressed".

I perform graphical and descriptive analyses to illustrate patterns of geolocation relative to ground-truth ZIP Code properties. I also use linear regression with ordinary least squares to study multivariate interactions among ZIP Code properties, geolocation accuracy, and ZIP Code identifications. In these analyses, I take the logarithm of geolocation error (plus 0.1), density, and area, since each raw distributions is severely skewed.

## 8.3. Results

### 8.3.1. General observations

Compared to ground-truth ZIP Codes, MaxMind is more likely to place users into a relatively few popular ZIP Codes (Figure 16). There are 13,800 unique ground-truth ZIP Codes and 11,600 unique MaxMind ZIP Codes.

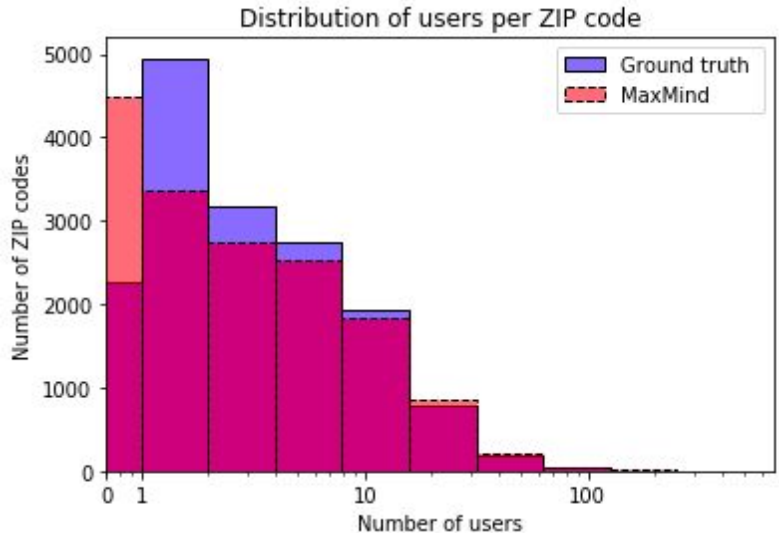

**Figure 16**: **Histogram of users per ZIP Code.** A histogram showing the distribution of user locations according to ground-truth and MaxMind data; the x-axis is logarithmic, and 0 is prepended. The purple region is the overlap of blue and red. MaxMind is more likely to place users into a smaller number of popular ZIP Codes (towards the right of the graph), and there are more ZIP Codes with zero MaxMind identifications than zero ground-truth identifications (far left).

According to ground-truth ZIP Codes, seven of the top ten ZIP Codes with highest per-capita registration rates are in Massachusetts (Table 1). Logically enough, the top two ZIP Codes are those that cover MIT's campus, with a highest registration rate of 18.2 registrants per 1000 population.

According to MaxMind, only four of the top ten ZIP Codes are in Massachusetts. The top two are in Illinois - and in the top one, in Chicago, the number of geolocated registrants is a remarkable 12% of the total population. Some of these ZIP Codes have small populations, and so the per-capita registration rates are sensitive to noise.

**Table 1.** ZIP Codes with highest per-capita usage (displayed in registrants per 1000 population)

| *According to ground-truth* | *According to MaxMind* |
|---|---|
| 02143 4.4 (Somerville, MA) | 02138 6.6 (Cambridge, MA) |
| 02134 4.4 (Allston, MA) | 92036 7.4 (Julian, CA) |
| 06103 4.7 (Hartford, CT) | 77002 8.2 (Houston, TX) |
| 08553 4.9 (Rocky Hill, NJ) | 02109 8.5 (Boston, MA) |
| 02120 5.0 (Roxbury, MA) | 02142 9.6 (Cambridge, MA) |
| 02141 5.2 (Cambridge, MA) | 97204 10.6 (Portland, OR) |
| 02210 5.7 (Boston, MA) | 95113 11.3 (San Jose, CA) |
| 97204 8.7 (Portland, OR) | 02139 13.6 (Cambridge, MA) |
| 02139 9.4 (Cambridge, MA) | 61602 15.1 (Peoria, IL) |
| 02142 18.2 (Cambridge, MA | 60602 121.4 (Chicago, IL) |

In line with the results from Chapter 7 according to ground-truth data users are far more likely to come from more prosperous areas. 41% are located in a Prosperous ZIP Code, 23% in Comfortable, 16% in Mid-tier, 12% in At risk, and 7% in Distressed (Figure 17).

One interesting question is whether the sample of users who provided parseable mailing addresses is representative of the general user population. Since my analyses rely on comparing two measurements for the same set of users, representativeness is not essential to demonstrate the existence of bias, but might affect its magnitude. I compare the distribution according to MaxMind of this sample with that of the much larger set from Chapter 7. The general patterns are similar (Figure 18), but this sample underrepresents users whom MaxMind geolocates to more prosperous ZIP Codes. This is consistent with Hansen and Reich's finding (Hansen and Reich 2015a) that users from higher socioeconomic status are less likely to provide mailing addresses.

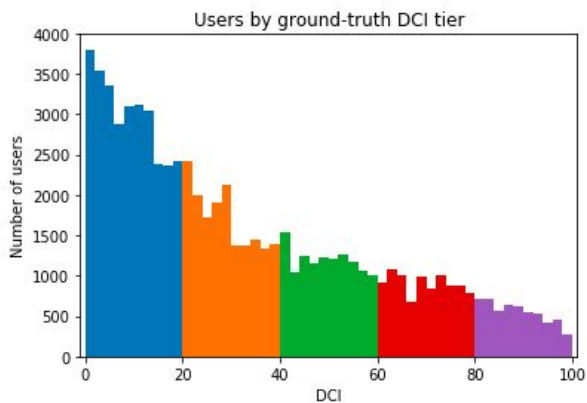
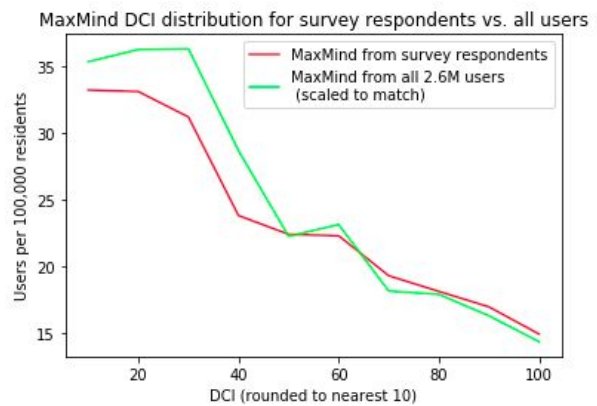

**Figure 17: Histogram of users by DCI.** Shows different DCI tiers in different colors (Prosperous at left, Distressed at right).

**Figure 18: Per-capita registrations by DCI.** Calculated by dividing the number of users whom MaxMind places in ZIP Codes of a particular sub-tier by the ZIP Codes' total population. The red line shows the registrants studied in this chapter, who provided mailing addresses; the green line shows the larger population studied in Chapter 7 (with numbers scaled to the same total number of users.)

I note that population, area, density, and DCI are reasonably independent metrics: of course population and area are each individually strongly associated with density, but otherwise correlations among all four are weak, whether we consider the raw measurement or its logarithm (Table 2).

**Table 2.** Correlations coefficients among ground-truth ZIP Code properties (calculated across users)

|             | Pop. | Log pop. | Area  | Log area | Density | Log density | DCI   |
|-------------|------|----------|-------|----------|---------|-------------|-------|
| Pop.        | 1    | 0.88     | -0.07 | -0.01    | 0.28    | 0.42        | 0.02  |
| Log pop.    |      | 1        | -0.09 | -0.03    | 0.22    | 0.49        | -0.03 |
| Area        |      |          | 1     | 0.48     | -0.13   | -0.47       | 0.1   |
| Log area    |      |          |       | 1        | -0.58   | -0.89       | 0     |
| Density     |      |          |       |          | 1       | 0.61        | 0.09  |
| Log density |      |          |       |          |         | 1           | -0.01 |
| DCI         |      |          |       |          |         |             | 1     |

Although centroid distance is defined and calculated differently than boundary distance, they are very strongly correlated (r = .99). There are a few cases where parsing or geocoding fails in which they differ substantially (Figure 19). I repeat analyses with centroid distance rather than boundary distance and find similar patterns of results, although sometimes with different magnitudes.

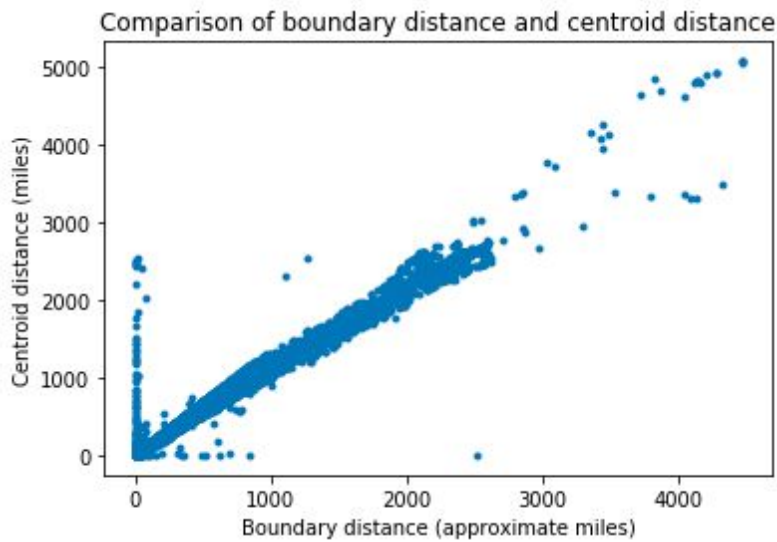

**Figure 19: Comparison of boundary distance and centroid distance.** A scatterplot showing boundary distance and centroid distance for the ~69,000 users where both are defined. Note that the vast majority of points are clumped in the bottom left.

## 8.3.2. Registration patterns

According to both ground-truth and MaxMind identifications, the vast majority of users come from ZIP Codes with greater population, smaller area, greater density, and lower DCI (Figure 20).

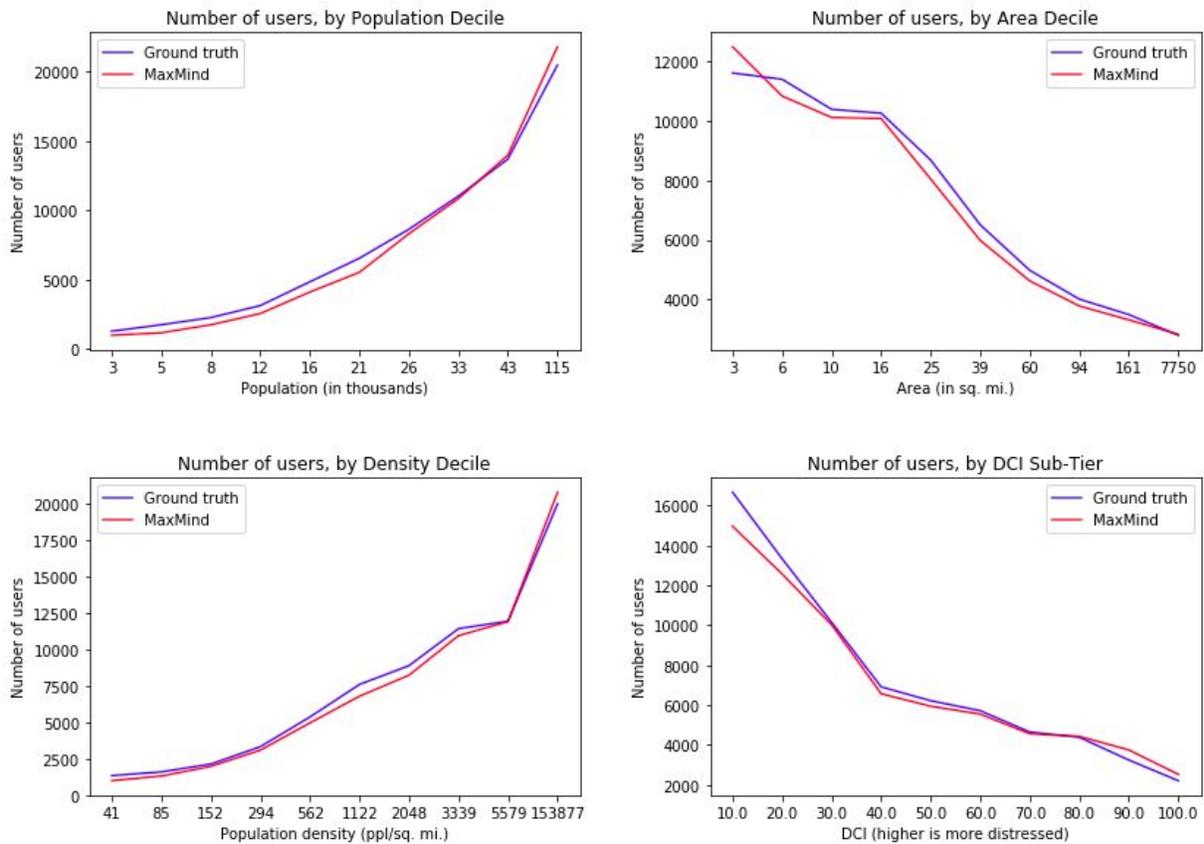

**Figure 20: Number of users by ZIP Code profile.** Number of users by level of population, area, density, and DCI, according to ground-truth and MaxMind. The x-axis labels represent the upper end of each bucket: for example, the first decile of population includes ZIP Codes with population under 3,000, and the tenth decile includes ZIP Codes with population under 115,000.

The patterns are somewhat more complex after adjusting for the total population of each ZIP Code level (Figure 21). Per-capita registration rates decrease with area and DCI. They also generally increase with population and density, after the three smallest deciles. The left portions of the graphs might be a statistical fluke: I calculate per-capita registration rates by dividing by a ZIP Code level's total population, so with smaller populations the calculation is more sensitive to noise.

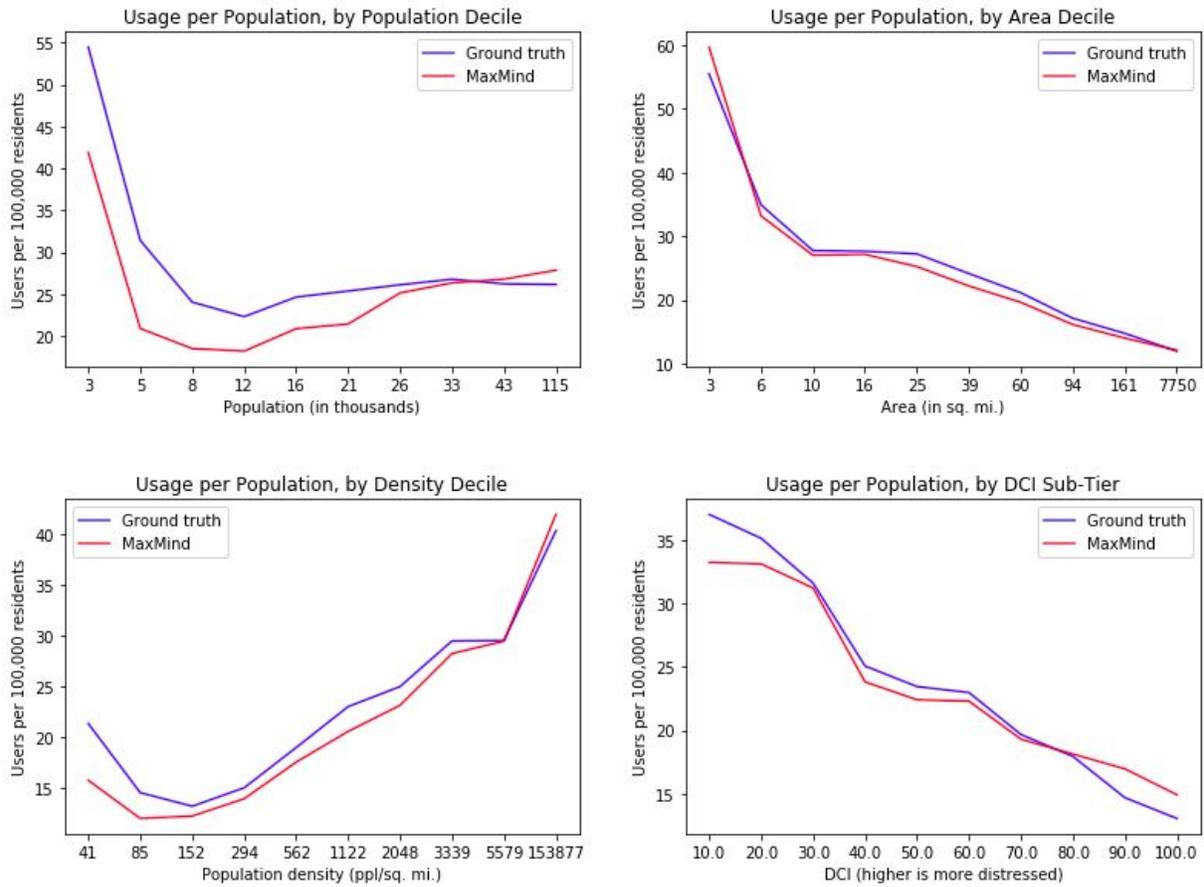

**Figure 21: Per-capita usage by ZIP Code profile.** Per-capita usage by level of population, area, density, and DCI, according to ground-truth and MaxMind. Again, the x-axis labels represent the upper end of each bucket.

### 8.3.3. Geolocation error

When studying geolocation accuracy, there are two broad considerations: the probability that MaxMind identifies the correct ZIP Code, and the magnitude of the error when it does not.

Overall, the MaxMind ZIP Code matches the ground-truth one for 18% of users, half the advertised number (6.3). The probability of an exact match increases with ground-truth population and ground-truth area (Table 3). The latter makes sense intuitively: landing in the correct ZIP Code seems easier if that area is physically large. The exact-match probability is highest for middle densities, and is relatively stable across DCI tiers.

Table 3. Geolocation accuracy by ground-truth ZIP Code properties
(1) Upper end of ZIP Code level
(2) Among users with ground-truth ZIP Code in this ZIP Code level, percentage where MaxMind identifies the exact ground-truth ZIP Code
(3) Among users with ground-truth ZIP Code in this level where MaxMind does not identify the exact ZIP Code, exponentiated mean of logarithm of geolocation error (measured as boundary distance)
(4) Among same users as (3), exponentiated standard deviation of logarithm of boundary distance

|  | (1) ZIP Code level: less than | (2) Percent exact-ZIP match | (3) Average geolocation error (approximate miles) | (4) Spread of geolocation error (approximate miles) |
| --- | --- | --- | --- | --- |
| Population (thousands) | 115 | 23.0 | 8.3 | 10.2 |
|  | 33 | 17.7 | 7.5 | 11.0 |
|  | 21 | 13.7 | 7.6 | 11.0 |
|  | 12 | 10.6 | 9.8 | 10.0 |
|  | 5 | 7.6 | 13.5 | 10.1 |
| Area (sq. mi.) | 7750 | 27.3 | 25.4 | 9.1 |
|  | 94 | 22.0 | 15.1 | 9.4 |
|  | 39 | 20.3 | 10.6 | 9.1 |
|  | 16 | 17.3 | 7.6 | 9.7 |
|  | 6 | 15.0 | 4.7 | 11.5 |
| Population density (people/sq. mi.) | 153877 | 16.9 | 5.5 | 11.1 |
|  | 3339 | 19.0 | 8.4 | 9.6 |
|  | 1122 | 21.5 | 12.8 | 9.5 |
|  | 294 | 21.1 | 18.3 | 8.8 |
|  | 85 | 17.2 | 29.8 | 7.8 |
| DCI | 100 | 18.2 | 9.1 | 11.4 |
|  | 80 | 18.9 | 8.3 | 11.5 |
|  | 60 | 19.3 | 8.6 | 11.1 |
|  | 40 | 18.2 | 8.1 | 11.1 |
|  | 20 | 18.6 | 8.0 | 9.6 |

When errors do occur, on average the errors decrease with population, decrease with population density, increase with DCI, and increase with area - so MaxMind is more likely to correctly geolocate a user from a bigger-area ZIP Code, but also to make a larger error if it fails.

The averages do not tell the whole story; it is useful to look at visual depictions of geolocation error in Figure 22 and Figure 23. Note that the distributions are strongly skewed right, and Figure 22 uses a logarithmic scale. The distributions of geolocation error shift to the right (larger errors) as density and population decrease. As area increases, the portion of zero-error identifications increases but the rest of the distribution moves right. As DCI tier increases, the left part of the error distribution remains similar but the right part spreads out more: the size of the biggest errors increases with DCI tier.

Figure 23 shows a different view. Here, we see the 10th, 20th, ..., 90th percentiles of geolocation error for particular groups of users. The 10th percentile is almost always zero, indicating that MaxMind matches least 10% of users to the exact ground-truth ZIP Code. But the higher percentiles change. In particular, for DCI the lower percentiles stay near-constant as DCI increases, but the higher percentiles move right: in more distressed regions, MaxMind makes greater errors at the tail of the distribution.

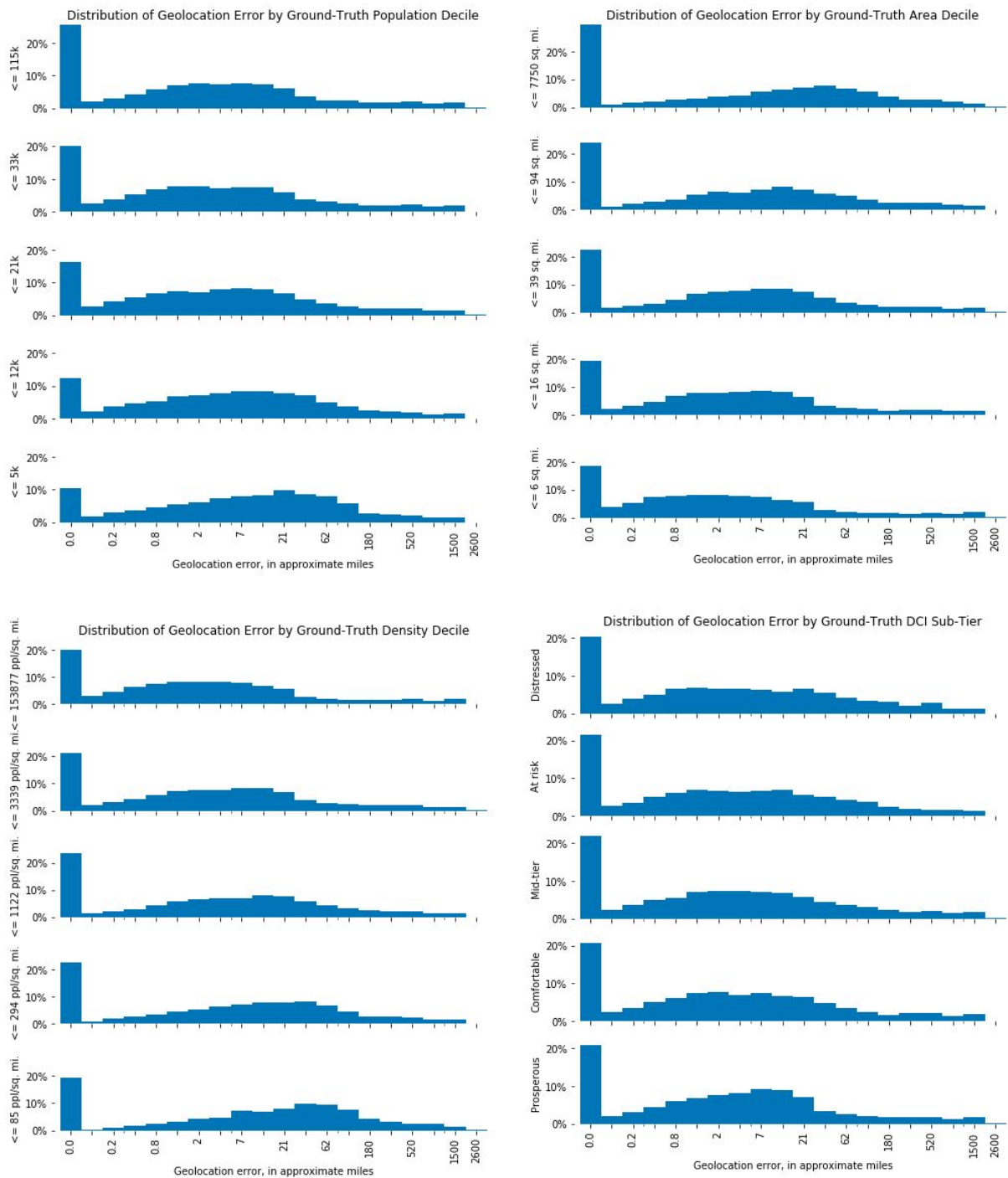

**Figure 22: Histograms of geolocation error.** Geolocation error, on a log scale (the logarithm was taken after adding a constant), across ZIP Code properties. The bar at the left of each chart includes the users whom MaxMind geolocated to the exact ground-truth ZIP Code, yielding a geolocation error of 0.

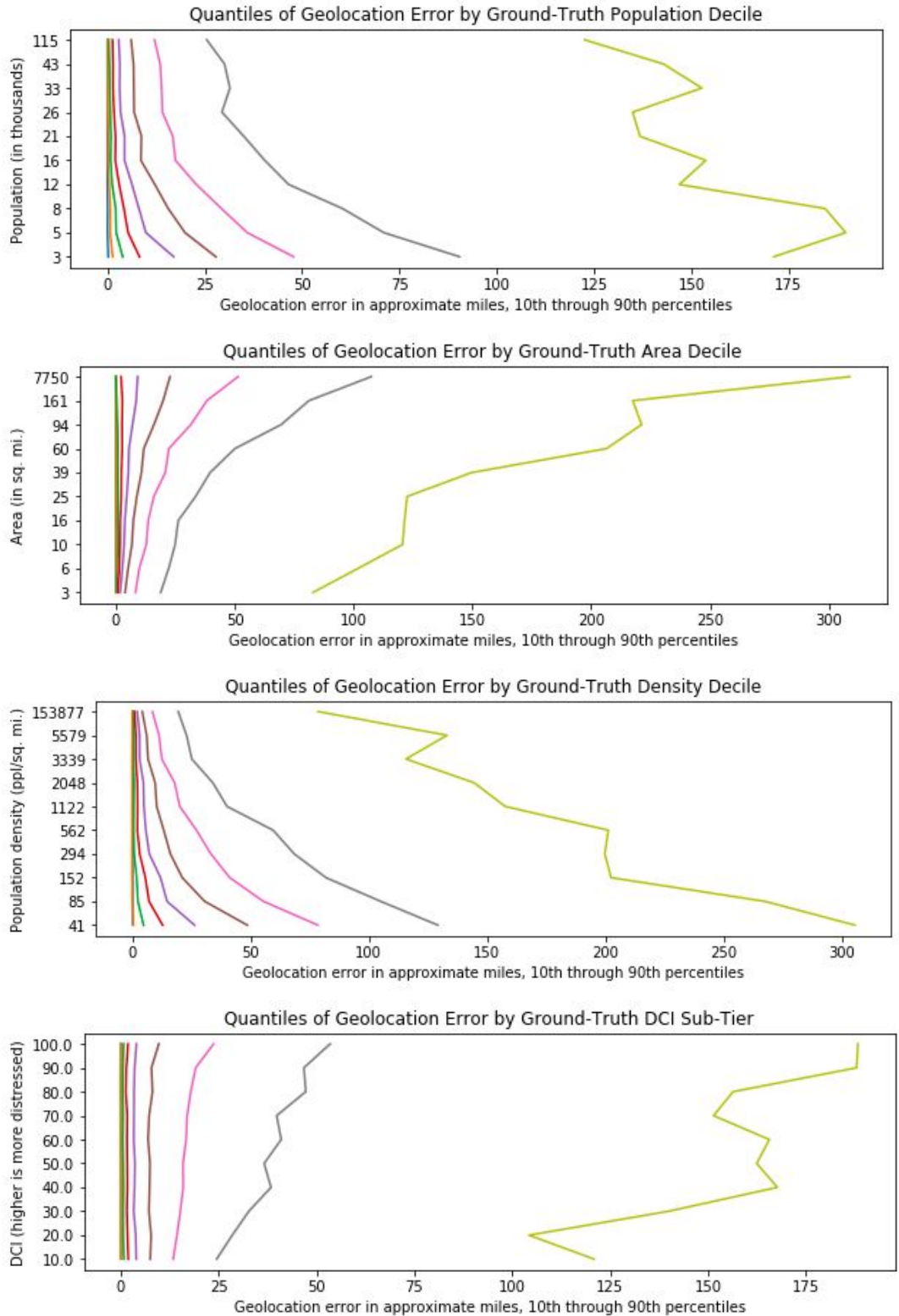

**Figure 23: Quantiles of geolocation error on a linear scale.** For each ZIP Code level, the graph shows the geolocation error at the 10th, 20th, 30th, ..., 90th percentile. For example, the last row of the last graph show users whose ground-truth ZIP Codes have DCI less than 10. The dark gray curve indicates that among these users, the 80th percentile of geolocation error is about 25 mi. The yellow-green shows that the 90th percentile is about 120 mi.

I use multivariate linear regression to quantify the relationships between ZIP Code factors and the probability of correctly identifying a ZIP Code, as well as with the size of geolocation error when the ZIP Code is incorrect.

An increase of 1000 population is associated with a 0.23 percentage point change in the probability of correctly identifying a ZIP Code and -0.21% change in geolocation error. A 1% increase in area is associated with a 0.03 percentage point change in correct identification probability and a 0.35% change in geolocation error. A 1-point increase in DCI is associated with near-zero change in correct identification probability and 0.22% change in geolocation error.

Except for the association between DCI and correct identification probability, all of these relationships are significant at the p = 0.001 level. Because the assumptions for ordinary least squares are not met, these numbers should be interpreted only as rough indicators. Appendices 11.3.1 and 11.3.2 contain full regression results.

### 8.3.4. ZIP Code property identification

A user's chances of being geolocated to a ZIP Code of roughly correct population, area, density, and DCI depends on his ground-truth ZIP Code (Figure 24). In each case, MaxMind geolocations are disproportionately drawn to one end of the distribution, and users from that favored tier are much more likely to be correctly identified.

For example, across all population quintiles except the second-highest, a user's geolocated ZIP Code is more likely to be in the highest population quintile than any other. A lowest-quintile user has a 20% chance of being correctly identified; a highest-quintile user has a 68% chance.

The trend is less dramatic for area, because area identifications are consistently more successful. Misidentifications are biased towards small areas. A lowest-quintile user has a 64% chance of correct identification, while a highest-quintile user has a 49% chance.

For density, higher tiers have the advantage. A lowest-quintile user has a 39% chance of correct identification, while a highest-quintile user has a 64% chance.

Geolocation is more successful for lower DCIs. A user from a Prosperous ZIP Code has a 59% chance of being geolocated to a Prosperous ZIP Code, while a user with a Distressed ground-truth ZIP Code has a 38% chance of being geolocated to a Distressed ZIP Code. Prosperous is the most common "wrong answer" across tiers.

I also quantify this economic bias by looking at the difference between the DCI of a user's ground-truth ZIP Code and her MaxMind-identified ZIP Code (Figure 25). Moving across the tiers from Prosperous to Distressed tiers, the median absolute differences are 8.6, 13.8, 19.1, 20.4, 25.1; the root-mean-squared differences are 28.8, 24.5, 26.2, 34.3, 43.6.

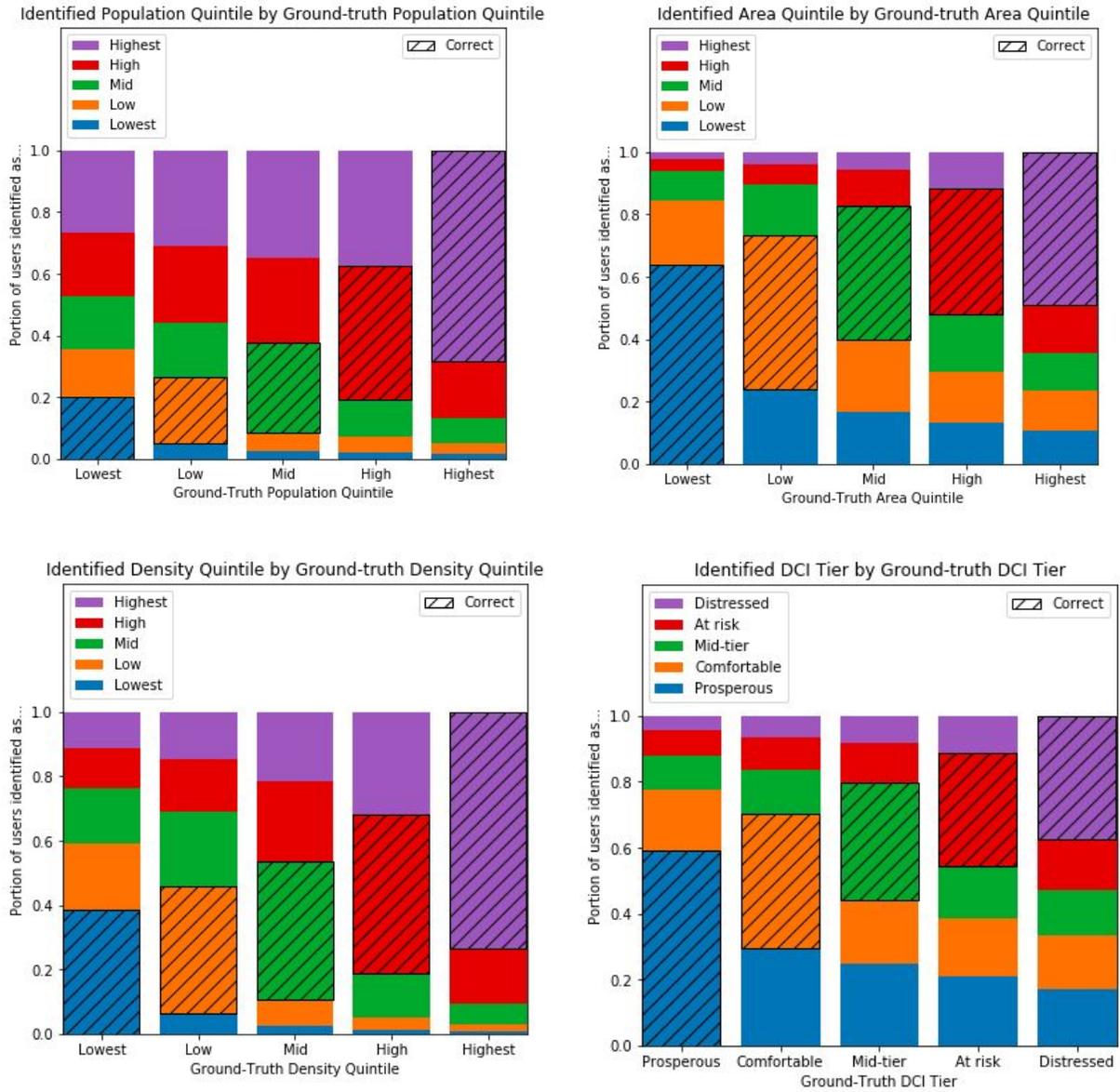

**Figure 24: MaxMind ZIP Code property by ground-truth ZIP Code property.** In each graph, each column represents the distribution of IP geolocation for users from a particular ground-truth ZIP Code tier. For example, the leftmost column of the last graph shows that among users with Prosperous ground-truth ZIP Codes, 59% are geolocated to Prosperous ZIP Codes (blue), 18% to Comfortable (orange), 11% to Mid-tier (green), 8% to At risk (red), and 4% to Distressed (purple).

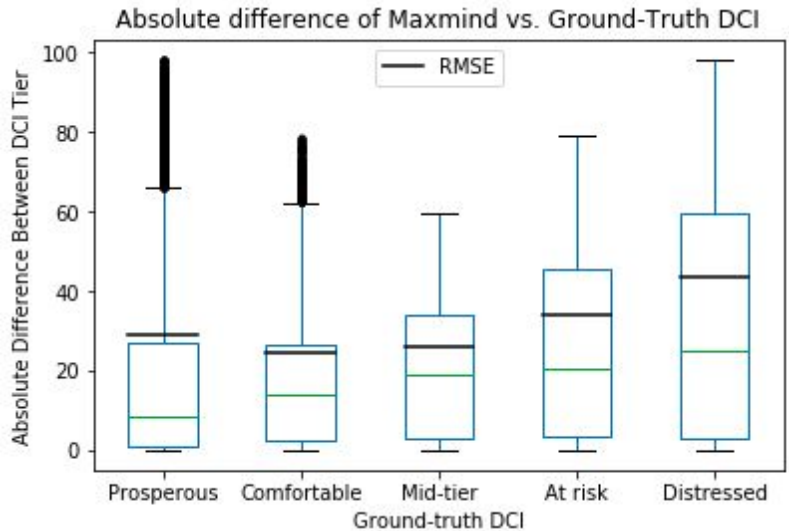

**Figure 25: Difference between MaxMind and ground-truth DCI.** Boxplot showing distribution of absolute value of difference between DCI of ground-truth and MaxMind ZIP Codes, by users' DCI tier. For each tier, the thick horizontal black line shows the root-mean-square error, the green line the median difference, the blue lines the first and third quartiles, the black whiskers the points 1.5 times the interquartile range away from the outer quartiles, and the black points outliers. Note that the absolute difference is by definition capped at 80 for the Comfortable and At risk tiers (DCI 20-40 and 60-80), and 60 for Mid-tier (DCI 40-60).

### 8.3.5. Overall DCI distributions

Studying the distribution of all geolocations rather than the probability distribution for individuals, I find that using IP geolocation *underestimates* the regressive pattern in edX usage (Figure 26). According to MaxMind geolocations, there are 33.2 users per 100,000 population in Prosperous areas and 16.1 in Distressed areas. According to ground-truth identifications, there are 36.2 in Prosperous areas and 14.0 in Distressed areas - a gap 30% bigger. Regressions showed that, after controlling for a ZIP Code's population and area, an increase of 10 in DCI is associated with 0.47 fewer users according to ground-truth counts, and 0.35 fewer users according to MaxMind counts (p < 0.001; standard errors of 0.02 and 0.03, respectively). Full results are in Appendix 11.3.3. Looking at the overall distribution (Figure 27), we see that the MaxMind distribution is less heavily skewed right than the ground-truth one, with fewer users from more prosperous areas and more from more distressed areas.

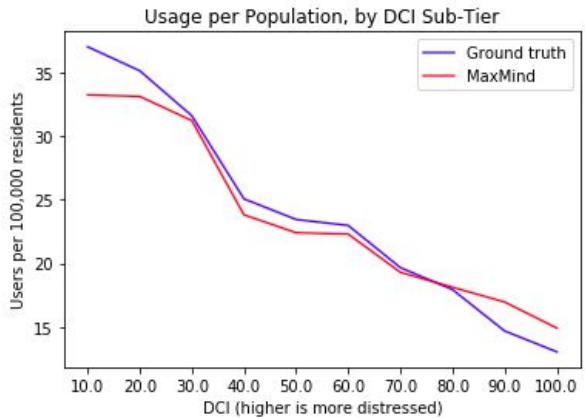 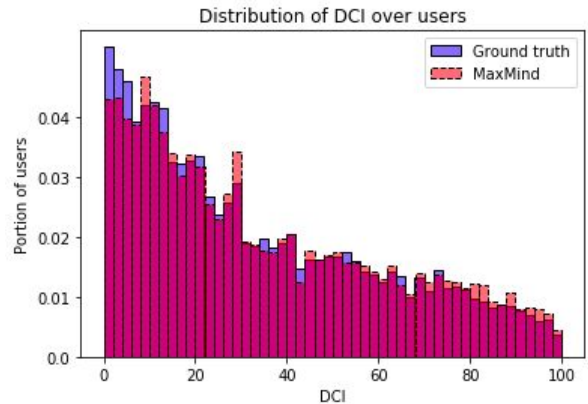

**Figure 26: Per-capita usage by DCI sub-tier.** According to ground-truth and MaxMind identifications; repeated from Figure 21.

**Figure 27: Histogram showing overall DCI distributions.** Blue, according to ground-truth; red, according to MaxMind identifications. The large purple area is the overlap of the red and the blue. Note that the ground-truth distribution is more skewed towards lower DCI than the MaxMind distribution.

This is an interesting dual effect in the patterns of DCI identifications. On the one hand, individual geolocations are biased towards more prosperous tiers: if we choose a user with equal probability from one of the five tiers, MaxMind is much more likely to identify her tier correctly if she is from a more prosperous area, and is much more likely to err on the side of putting her in a more prosperous than a more distressed area. On the other hand, in total the MaxMind distribution is shifted more towards more distressed areas. Figure 28 helps explain this seeming paradox: there are far fewer users from distressed areas than from prosperous ones, where MaxMind can err only in the too-distressed direction. Although MaxMind is less likely to err for an individual prosperous user than an individual distressed user, in total it makes more errors for prosperous users than distressed ones. 46.4% of users are placed in the correct DCI tier, 28.4% in one that is more distressed, and 25.3% in one that is less distressed.

Having no reason to think that geolocation is systematically biased differently on my dataset than on the general population, I suspect that the first effect is general: MaxMind geolocation (and likely its competitors) will more accurately identify an economic tier for users from more prosperous areas, and disproportionately mislocate users from distressed areas to prosperous ones. The second one is specific to my dataset, in which users from more prosperous areas are far more common. If instead users' ground-truth was uniformly distributed across tiers, the MaxMind-given distribution would overrepresent prosperous areas. If instead users from distressed areas were overrepresented, the MaxMind-given distribution would reduce this bias: the noisy measurement obscures the signal, in both directions.

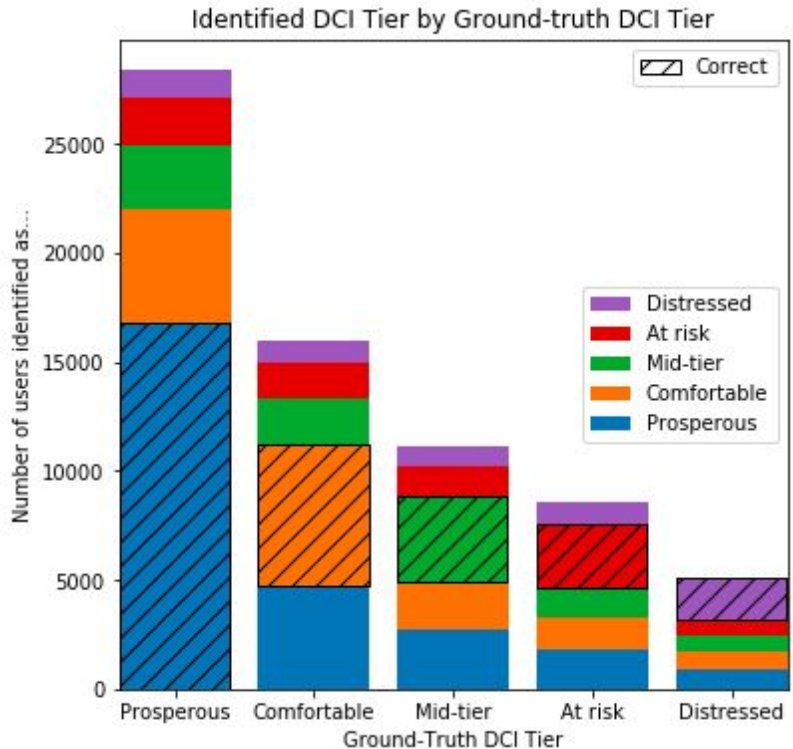

**Figure 28: Identified DCI tiers, scaled.** DCI tier of MaxMind ZIP Code by DCI tier of ground-truth ZIP Code, in absolute numbers. Each column shows the number of people from a particular ground-truth DCI tier whom MaxMind placed in each DCI tier. This is a rescaled version of Figure 24.

## 8.4. Discussion

Using IP geolocation to analyze users' economic patterns produces biased results and penalizes users from more economically distressed areas. Users from less dense and more distressed areas experience larger geolocation errors, and are more likely to be geolocated to ZIP Codes that doesn't resemble their true ones. Geolocated ZIP Codes also underestimate the regressive patterns in MOOC usage: by ground-truth data, users from wealthy areas dominate MOOCs even more than the results from IP geolocation indicate. The results suggest that researchers must use geolocation with great caution not only because it is often inaccurate, as the networking literature suggests, but also because it can introduce bias to the data.

All the work above comes with caveats. The biggest is that I refer to users' mailing addresses as their "ground-truth" locations, but this is probably not quite right. Users might put in their work addresses but access edX from home, or vice versa (tellingly, MIT's main address often appears in the data); college students might enter their families' addresses but access courses from campus; some P.O. boxes might be located far from users' residences; users might move or travel while taking a course; some might fabricate false addresses. There could even be systematic biases of reported addresses coming from more or less prosperous places than physical locations - for example, if offices tend to be located in more distressed regions than homes. Nonetheless, I think it is fair to say that users' mailing addresses are a substantially *more* accurate indicator of their locations than their IP geolocation, which even by MaxMind's assessment is correct at the ZIP Code level scarcely a third of the time.

I cannot blame all the geolocation errors on the accuracy of MaxMind's database. Some errors probably happen because users, or their networks, hide their devices' IP addresses with technologies like proxies or network address translation. For a researcher who needs accurate geolocations, it doesn't much matter whether an error happens because MaxMind incorrectly ties an IP address to a location or because MaxMind is given the wrong IP address - but the database's accuracy cannot be held responsible in the latter case.

There are some smaller caveats as well: although I spot-check address parsing and geocoding, certainly both processes make some mistakes. It is possible that the supplementary datasets used have systematically biased errors or omissions. The different datasets used are not perfectly aligned in time or space.

# 9. Discussion

I summarize my key findings in 9.1 and discuss their implications in 9.2. I discuss limitations of this work in 9.3 and offer suggestions for future work in 9.4.

## 9.1. Summary of key findings

In Chapter 7, I use IP geolocation with MaxMind to substantiate the results of Hansen and Reich 2015b: according to IP geolocation, HarvardX and MITx users disproportionately come from ZIP Codes of greater economic prosperity and population density. Additionally, I find that per-ZIP Code registration rates across courses are largely unidimensional; studying registration rates across individual courses gives little more information than considering all courses together.

In Chapter 8, I examine the accuracy of IP geolocation by comparing geolocated ZIP Codes to "ground-truth" self-provided mailing addresses. I find that IP geolocation misrepresents users from disadvantaged ZIP Codes in several ways: MaxMind places users from more distressed, less dense areas to ZIP Codes farther from their ground-truth ones. MaxMind is also less likely to place these users in ZIP Codes that are economically similar to their ground-truth ones, instead often placing them in more dense, more prosperous areas. Furthermore, IP geolocation underestimates the lack of users from distressed areas: the gap between per-capita usage in Prosperous and Distressed ZIP Codes is 30% greater according to ground-truth ZIP Codes than MaxMind ones.

## 9.2. Implications

Using either mailing addresses or IP geolocation, users from disadvantaged areas are underrepresented among MOOC participants. This points in the same direction as previous research: MOOCs do not seem to be democratizing education, but rather providing more resources to people who already likely have more access to wealth, employment, and education. To reverse this pattern, MOOC developers would likely have to invest great resources into targeting less advantaged areas - or they would have to accept that MOOCs are not the answer to revolutionizing education.

Researchers must use great care when identifying MOOC users' locations with fine-grained IP geolocation. IP geolocation makes greater errors for users from more distressed areas and biases results towards more prosperous areas. As a noisy measurement, it can also obscure patterns in overall distributions - in this case, underestimating the economically regressive pattern in usage.

I suspect that these biases have affected existing research. For example, Alcorn, Christensen, and Kapur 2015 geolocated IP address to cities, finding that "few rural residents of India are enrolling in MOOCs" - but if geolocation biases are similar in India as in the U.S., it is likely that many existing rural MOOC users were misidentified as being urban. I also think it likely that similar problems apply to other IP geolocation databases, as prior research has found that they function similarly to MaxMind (Shavitt and Zilberman 2011), and to non-MOOC social science research that uses IP geolocation.

IP geolocation may remain a useful tool for research. Lacking survey responses, IP addresses may be the best information available for suggesting users' locations. In my experience, it is also faster and cheaper to approximate locations by inputting IP addresses to a database than by parsing and geocoding freeform mailing addresses. However, at a minimum researchers should treat results of IP geolocation with skepticism. They should acknowledge that IP geolocations are at best approximations, and attempt to corroborate IP findings with alternatives such as user-provided locations. They should be particularly wary when studying economic patterns.

As discussed in 5.3, IP geolocation is used in many applications outside of research. This evidence of economic bias in its results raises a worrying possibility that it introduces unfair biases in advertising, pricing, and even high-stakes law enforcement scenarios.

I conjecture that there are two key reasons that more-advantaged areas experience more accurate geolocation: first, that these areas have more developed Internet infrastructure (Whitacre and Mills 2007); second, that ISPs and geolocation database providers have greater financial incentives to provide accurate information for areas where there are more and wealthier users. Large-scale investments would be needed to remedy the first problem, but pressure from researchers and companies could help the second.

## 9.3. Limitations

The datasets I use contained errors and are imperfectly aligned, as discussed in 6.7 and 8.4.

Throughout, I study the economic profiles of users' ZIP Codes, not of users themselves. It stands to reason that these are related: since most users are from economically prosperous areas, probably most users are economically prosperous. But the relationship is imperfect; some evidence shows that neighborhood socioeconomic status is a biased proxy for the individual (5.6), and it is possible that MOOC users are consistently wealthier or poorer than their neighbors.

I cannot verify the generalizability of my results. I study MOOCs offered on a popular platform by two of the world's most elite universities. It is possible that MOOCs offered by other institutions or on other platforms have different patterns of usage. It may also be that the IP geolocation biases I find are specific to MaxMind (not its competitors), or to edX registrants (not other groups of Internet users).

Geolocation errors are not always the result of inaccuracy in MaxMind's database: the IP addresses that edX records need not always correspond to users' physical devices (as explained in 5.4), nor mailing addresses to physical locations (8.4). My results provide strong reason for researchers to take great care in using IP geolocation, but only weak evidence that the MaxMind dataset itself is systematically biased or less accurate than claimed.

## 9.4. Future work

Future work could extend in several different directions.

In studying the relationships between MOOC usage and local economic data, it would be interesting to look at more detailed patterns in course engagement - metrics like completion, certification, video watching, problem solving, forum participation, and grades, rather than just registration. For example, if it emerged that people from less-prosperous ZIP Codes learned more by completing MOOCs despite being less likely to register, this could suggest that MOOCs still have a promising role in effecting economic change. It would also be interesting to break down the results over different years, tracking changes in MOOC users' economic distributions over time.

In studying IP geolocation bias, one goal would be to verify these results using known-correct data. A perfect gold standard dataset is very difficult to achieve. An approximation might include an Internet Service Provider's detailed directory of IP address assignments alongside user-reported current ZIP Codes, financial information, and demographics - if users have an incentive to report completely and truthfully.

More practically, one possibility for supplementing the dataset might be to use ZIP Codes associated with credit cards, for those who pay for certification. These data would not be perfect - the card might belong to a relative's friend or employer, or the ZIP Code to a former residence - but may be useful for cross-checking self-reported mailing addresses.

A larger group of research questions centers around extending these results to a broader context: does IP geolocation introduce economic bias when studying a representative sample of edX users, rather than possibly-nonrepresentative survey respondents? In MOOCs beyond edX? On the level of the individual user rather than the ZIP Code? Alternatively, at the level of the state, region, or country? Using competitors to MaxMind? In non-MOOC research? In commercial and governmental applications?

Another goal would be to improve DCI prediction by taking advantage of other user data, not just IP address. It might be possible to use machine learning methods to more accurately predict user backgrounds and correct for some of the biases and mistakes in IP geolocation. I explored using course engagement data and some self-reported demographics without success, but an effective approach may exist.

# 11. Appendices

## 11.1. Factor analysis correlations for ZIP Codes

Table 4 displays correlations between ZIP Code MOOC engagement variables, ZIP Code socioeconomic variables, and first and second factor scores, color-coded by magnitude of correlation coefficient (more positive is more blue, more negative is more red).

## 11.2. Factor analysis correlations for Courses

Table 5 displays correlations between course engagement variables, first-factor loadings, first-factor uniqueness, and item-test correlations (using registration data), color-coded by magnitude of correlation coefficient (more positive is more blue, more negative is more red).

**Table 4.** Correlations between ZIP Code variables; continues on next page.

| | Logarithm of population | Population density | Percent minority | Poverty rate | Change in establishments | Population | Change in employment | DCI | Median income ratio | Housing vacancies | Unemployment | Adults without HS degrees | Factor 1, from registrations | Factor 2, from registrations | Factor 1, from completions | Factor 2, from completions |
|---|---|---|---|---|---|---|---|---|---|---|---|---|---|---|---|---|
| Logarithm of population | 1.00 | 0.42 | 0.43 | 0.10 | 0.12 | 0.87 | 0.07 | -0.03 | 0.00 | -0.13 | -0.07 | 0.16 | -0.08 | -0.04 | -0.11 | -0.05 |
| Population density | | 1.00 | 0.54 | 0.18 | 0.08 | 0.41 | 0.00 | 0.02 | -0.04 | -0.01 | -0.09 | 0.18 | 0.14 | 0.04 | 0.11 | 0.06 |
| Percent minority | | | 1.00 | 0.48 | 0.11 | 0.47 | 0.02 | 0.36 | -0.29 | 0.22 | 0.27 | 0.60 | 0.04 | 0.03 | 0.00 | 0.00 |
| Poverty rate | | | | 1.00 | -0.13 | 0.13 | -0.12 | 0.82 | -0.72 | 0.59 | 0.62 | 0.66 | 0.03 | 0.01 | 0.01 | 0.00 |
| Change in establishments | | | | | 1.00 | 0.14 | 0.41 | -0.34 | 0.20 | -0.16 | -0.11 | -0.09 | 0.00 | 0.01 | -0.01 | 0.00 |
| Population | | | | | | 1.00 | 0.07 | 0.01 | -0.05 | -0.12 | -0.03 | 0.25 | -0.05 | -0.03 | -0.08 | -0.04 |
| Change in employment | | | | | | | 1.00 | -0.34 | 0.18 | -0.15 | -0.10 | -0.08 | 0.01 | 0.01 | 0.00 | 0.01 |
| DCI | | | | | | | | 1.00 | -0.78 | 0.68 | 0.71 | 0.70 | -0.05 | -0.03 | -0.05 | -0.03 |
| Median income ratio | | | | | | | | | 1.00 | -0.48 | -0.50 | -0.60 | 0.06 | 0.04 | 0.06 | 0.04 |
| Housing vacancies | | | | | | | | | | 1.00 | 0.51 | 0.34 | 0.01 | -0.02 | 0.00 | 0.00 |
| Unemployment | | | | | | | | | | | 1.00 | 0.52 | -0.04 | -0.01 | -0.05 | -0.03 |
| Adults without HS degrees | | | | | | | | | | | | 1.00 | -0.07 | -0.03 | -0.09 | -0.05 |
| Factor 1, from registrations | | | | | | | | | | | | | 1.00 | 0.66 | 0.86 | 0.53 |
| Factor 2, from registrations | | | | | | | | | | | | | | 1.00 | 0.41 | 0.29 |
| Factor 1, from completions | | | | | | | | | | | | | | | 1.00 | 0.74 |
| Factor 2, from completions | | | | | | | | | | | | | | | | 1.00 |

**Table 4**, continued. Correlations between ZIP Code variables

| Factor 1, from certifications | Factor 2, from certifications | Certifications per population | Completions per population | Registrations per population | Log of registration factor 1 | Log of registrations factor 2 | Log of completion factor 1 | Log of completion factor 2 | Log of certification factor 1 | Log of certification factor 2 | Log of registrations per million | Log of certification per million | Log of completions per million | Completion rate | Log of completion rate | Certification rate | Log of certification rate |
|---|---|---|---|---|---|---|---|---|---|---|---|---|---|---|---|---|---|
| -0.11 | 0.02 | -0.21 | -0.20 | -0.08 | 0.08 | -0.04 | -0.16 | -0.05 | -0.16 | 0.02 | 0.08 | -0.34 | -0.32 | -0.46 | -0.53 | -0.45 | -0.53 |
| 0.12 | -0.05 | 0.10 | 0.09 | 0.14 | 0.35 | 0.04 | 0.16 | 0.07 | 0.16 | -0.05 | 0.36 | 0.09 | 0.10 | -0.25 | -0.30 | -0.24 | -0.30 |
| 0.00 | -0.02 | -0.03 | -0.03 | 0.03 | 0.06 | 0.03 | -0.08 | 0.01 | -0.08 | -0.02 | 0.05 | -0.16 | -0.15 | -0.18 | -0.23 | -0.18 | -0.22 |
| 0.01 | 0.00 | 0.00 | 0.00 | 0.03 | -0.08 | 0.01 | -0.13 | 0.00 | -0.13 | -0.01 | -0.09 | -0.15 | -0.15 | -0.02 | -0.03 | -0.01 | -0.03 |
| -0.01 | -0.01 | -0.02 | -0.02 | 0.00 | 0.02 | 0.01 | -0.01 | 0.00 | -0.01 | -0.01 | 0.03 | -0.03 | -0.03 | -0.05 | -0.07 | -0.05 | -0.06 |
| -0.08 | 0.02 | -0.13 | -0.13 | -0.05 | 0.09 | -0.03 | -0.10 | -0.04 | -0.10 | 0.02 | 0.09 | -0.23 | -0.21 | -0.30 | -0.37 | -0.30 | -0.37 |
| 0.00 | -0.02 | -0.01 | -0.01 | 0.01 | 0.01 | 0.01 | -0.03 | 0.01 | -0.03 | -0.02 | 0.01 | -0.04 | -0.05 | -0.06 | -0.07 | -0.06 | -0.06 |
| -0.06 | 0.02 | -0.06 | -0.06 | -0.05 | -0.24 | -0.03 | -0.21 | -0.04 | -0.21 | 0.02 | -0.26 | -0.20 | -0.20 | 0.09 | 0.09 | 0.08 | 0.09 |
| 0.06 | -0.02 | 0.07 | 0.07 | 0.06 | 0.23 | 0.04 | 0.21 | 0.04 | 0.20 | -0.02 | 0.24 | 0.20 | 0.20 | -0.07 | -0.07 | -0.07 | -0.07 |
| 0.00 | 0.01 | 0.01 | 0.01 | 0.01 | -0.11 | -0.02 | -0.09 | -0.01 | -0.09 | 0.00 | -0.12 | -0.06 | -0.07 | 0.08 | 0.09 | 0.08 | 0.09 |
| -0.05 | 0.02 | -0.05 | -0.05 | -0.04 | -0.24 | -0.01 | -0.21 | -0.03 | -0.20 | 0.02 | -0.25 | -0.19 | -0.19 | 0.09 | 0.09 | 0.09 | 0.10 |
| -0.09 | 0.02 | -0.11 | -0.10 | -0.07 | -0.26 | -0.03 | -0.25 | -0.05 | -0.24 | 0.02 | -0.27 | -0.28 | -0.28 | 0.03 | 0.02 | 0.03 | 0.02 |
| 0.86 | -0.41 | 0.87 | 0.88 | 0.99 | 0.50 | 0.69 | 0.39 | 0.54 | 0.35 | -0.41 | 0.51 | 0.42 | 0.42 | -0.06 | -0.08 | -0.07 | -0.08 |
| 0.39 | -0.29 | 0.39 | 0.40 | 0.63 | 0.25 | 1.00 | 0.18 | 0.30 | 0.16 | -0.28 | 0.25 | 0.18 | 0.19 | -0.04 | -0.05 | -0.04 | -0.05 |
| 0.99 | -0.58 | 0.94 | 0.94 | 0.88 | 0.45 | 0.44 | 0.48 | 0.74 | 0.43 | -0.60 | 0.45 | 0.48 | 0.49 | 0.03 | 0.04 | 0.02 | 0.03 |
| 0.72 | -0.91 | 0.61 | 0.61 | 0.54 | 0.24 | 0.31 | 0.27 | 1.00 | 0.25 | -0.92 | 0.25 | 0.26 | 0.27 | 0.01 | 0.02 | 0.01 | 0.01 |
| 1.00 | -0.56 | 0.95 | 0.93 | 0.87 | 0.45 | 0.42 | 0.48 | 0.72 | 0.45 | -0.57 | 0.45 | 0.50 | 0.49 | 0.03 | 0.04 | 0.02 | 0.04 |
|  | 1.00 | -0.44 | -0.44 | -0.42 | -0.16 | -0.31 | -0.16 | -0.91 | -0.16 | 1.00 | -0.16 | -0.15 | -0.15 | 0.02 | 0.02 | 0.01 | 0.02 |
|  |  | 1.00 | 0.98 | 0.88 | 0.49 | 0.42 | 0.52 | 0.62 | 0.48 | -0.44 | 0.50 | 0.59 | 0.59 | 0.09 | 0.11 | 0.09 | 0.10 |
|  |  |  | 1.00 | 0.89 | 0.49 | 0.43 | 0.51 | 0.62 | 0.46 | -0.45 | 0.50 | 0.57 | 0.58 | 0.08 | 0.10 | 0.07 | 0.08 |
|  |  |  |  | 1.00 | 0.49 | 0.65 | 0.38 | 0.55 | 0.34 | -0.42 | 0.50 | 0.41 | 0.41 | -0.07 | -0.08 | -0.07 | -0.08 |
|  |  |  |  |  | 1.00 | 0.27 | 0.66 | 0.26 | 0.61 | -0.15 | 0.99 | 0.65 | 0.68 | -0.36 | -0.40 | -0.37 | -0.41 |
|  |  |  |  |  |  | 1.00 | 0.19 | 0.33 | 0.17 | -0.30 | 0.27 | 0.20 | 0.20 | -0.04 | -0.06 | -0.04 | -0.06 |
|  |  |  |  |  |  |  | 1.00 | 0.28 | 0.94 | -0.16 | 0.66 | 0.87 | 0.89 | 0.13 | 0.19 | 0.11 | 0.16 |
|  |  |  |  |  |  |  |  | 1.00 | 0.26 | -0.92 | 0.26 | 0.28 | 0.28 | 0.01 | 0.02 | 0.01 | 0.01 |
|  |  |  |  |  |  |  |  |  | 1.00 | -0.15 | 0.61 | 0.86 | 0.84 | 0.13 | 0.19 | 0.12 | 0.18 |
|  |  |  |  |  |  |  |  |  |  | 1.00 | -0.16 | -0.15 | -0.15 | 0.01 | 0.02 | 0.01 | 0.01 |
|  |  |  |  |  |  |  |  |  |  |  | 1.00 | 0.66 | 0.68 | -0.38 | -0.41 | -0.38 | -0.43 |
|  |  |  |  |  |  |  |  |  |  |  |  | 1.00 | 0.98 | 0.24 | 0.31 | 0.23 | 0.31 |
|  |  |  |  |  |  |  |  |  |  |  |  |  | 1.00 | 0.23 | 0.30 | 0.20 | 0.26 |
|  |  |  |  |  |  |  |  |  |  |  |  |  |  | 1.00 | 0.95 | 0.99 | 0.95 |
|  |  |  |  |  |  |  |  |  |  |  |  |  |  |  | 1.00 | 0.94 | 0.98 |
|  |  |  |  |  |  |  |  |  |  |  |  |  |  |  |  | 1.00 | 0.95 |
|  |  |  |  |  |  |  |  |  |  |  |  |  |  |  |  |  | 1.00 |

**Table 5.** Correlations among course variables.

| | Uniqueness from registrations | Factor loadings from registrations | Uniqueness from certifications | Factor loadings from certifications | Uniqueness from completions | Factor loadings from completions | Certifications | Completions | Registrations | Completion rate | Certification rate | Item-Test corr from registrations | Log of registrations | Log of completions | Log of certifications |
|---|---|---|---|---|---|---|---|---|---|---|---|---|---|---|---|
| Uniqueness from registrations | 1.00 | -0.98 | 0.40 | -0.46 | 0.39 | -0.44 | -0.40 | -0.42 | -0.32 | 0.07 | 0.01 | -0.97 | -0.72 | -0.56 | -0.55 |
| Factor loadings from registrations | -0.98 | 1.00 | -0.37 | 0.43 | -0.36 | 0.41 | 0.38 | 0.40 | 0.29 | -0.10 | -0.05 | 0.98 | 0.72 | 0.55 | 0.54 |
| Uniqueness from certifications | 0.40 | -0.37 | 1.00 | -0.95 | 0.90 | -0.86 | -0.66 | -0.64 | -0.31 | -0.21 | -0.36 | -0.41 | -0.44 | -0.53 | -0.58 |
| Factor loadings from certifications | -0.46 | 0.43 | -0.95 | 1.00 | -0.85 | 0.89 | 0.66 | 0.65 | 0.33 | 0.26 | 0.43 | 0.48 | 0.52 | 0.64 | 0.70 |
| Uniqueness from completions | 0.39 | -0.36 | 0.90 | -0.85 | 1.00 | -0.95 | -0.57 | -0.58 | -0.25 | -0.25 | -0.32 | -0.38 | -0.38 | -0.51 | -0.51 |
| Factor loadings from completions | -0.44 | 0.41 | -0.86 | 0.89 | -0.95 | 1.00 | 0.58 | 0.60 | 0.28 | 0.32 | 0.38 | 0.44 | 0.45 | 0.63 | 0.62 |
| Certifications | -0.40 | 0.38 | -0.66 | 0.66 | -0.57 | 0.58 | 1.00 | 0.97 | 0.34 | 0.28 | 0.45 | 0.45 | 0.61 | 0.71 | 0.74 |
| Completions | -0.42 | 0.40 | -0.64 | 0.65 | -0.58 | 0.60 | 0.97 | 1.00 | 0.33 | 0.37 | 0.44 | 0.46 | 0.62 | 0.75 | 0.75 |
| Registrations | -0.32 | 0.29 | -0.31 | 0.33 | -0.25 | 0.28 | 0.34 | 0.33 | 1.00 | -0.09 | -0.03 | 0.36 | 0.57 | 0.32 | 0.34 |
| Completion rate | 0.07 | -0.10 | -0.21 | 0.26 | -0.25 | 0.32 | 0.28 | 0.37 | -0.09 | 1.00 | 0.81 | -0.12 | -0.13 | 0.47 | 0.38 |
| Certification rate | 0.01 | -0.05 | -0.36 | 0.43 | -0.32 | 0.38 | 0.45 | 0.44 | -0.03 | 0.81 | 1.00 | -0.04 | -0.01 | 0.47 | 0.53 |
| Item-Test corr from registrations | -0.97 | 0.98 | -0.41 | 0.48 | -0.38 | 0.44 | 0.45 | 0.46 | 0.36 | -0.12 | -0.04 | 1.00 | 0.80 | 0.59 | 0.60 |
| Log of registrations | -0.72 | 0.72 | -0.44 | 0.52 | -0.38 | 0.45 | 0.61 | 0.62 | 0.57 | -0.13 | -0.01 | 0.80 | 1.00 | 0.72 | 0.75 |
| Log of completions | -0.56 | 0.55 | -0.53 | 0.64 | -0.51 | 0.63 | 0.71 | 0.75 | 0.32 | 0.47 | 0.47 | 0.59 | 0.72 | 1.00 | 0.94 |
| Log of certifications | -0.55 | 0.54 | -0.58 | 0.70 | -0.51 | 0.62 | 0.74 | 0.75 | 0.34 | 0.38 | 0.53 | 0.60 | 0.75 | 0.94 | 1.00 |

## 11.3. Regression results

This appendix displays the results of regressing:
- logarithm of geolocation errors, on users' ground-truth ZIP Code economic properties (11.3.1).
- probability of a user's MaxMind and ground-truth ZIP Codes matching, on ground-truth ZIP Code properties (11.3.2).
- Number of users in a ZIP Code according to MaxMind and ground-truth, on ZIP Code economic properties (11.3.3).

The variables used in each case are DCI (0 to 100), population in thousands of residents, and natural logarithm of area in square miles. I show results of the regressions when adding one variable at a time. I provide an interpretation of the full regression at the bottom of each table. The tables show coefficients, standard errors, t-scores, p-values, and the 2.5%-97.5% confidence interval.

### 11.3.1. Geolocation error, by user

I show results for boundary distance in approximate miles (including users with no geolocation error in Table 6 and excluding them in Table 7) and centroid distance in miles (including users with no geolocation error in Table 8 and excluding them in Table 9).

**Table 6. Regressions for logarithm of boundary distance**
```
==============================================================================
                 coef    std err          t      P>|t|      [0.025      0.975]
------------------------------------------------------------------------------
const          1.2755      0.017     77.303      0.000       1.243       1.308
dci            0.0010      0.000      2.561      0.010       0.000       0.002
==============================================================================
==============================================================================
                 coef    std err          t      P>|t|      [0.025      0.975]
------------------------------------------------------------------------------
const          1.6832      0.024     69.023      0.000       1.635       1.731
dci            0.0012      0.000      3.132      0.002       0.000       0.002
population    -0.0122      0.001    -22.636      0.000      -0.013      -0.011
==============================================================================
==============================================================================
                 coef    std err          t      P>|t|      [0.025      0.975]
------------------------------------------------------------------------------
const          1.3111      0.030     43.247      0.000       1.252       1.371
dci            0.0013      0.000      3.397      0.001       0.001       0.002
population    -0.0122      0.001    -22.658      0.000      -0.013      -0.011
ln_area        0.1475      0.007     20.549      0.000       0.133       0.162
==============================================================================
```

Interpretation: when considering all users, an increase of 1 in DCI is associated with a 0.13% change in boundary distance, an increase of 1000 population is associated with a -1.2% change in boundary distance, and a 1% increase in area is associated with a 0.15% change in boundary distance.

**Table 7. Regressions for logarithm of boundary distance, excluding zeros**
```
==============================================================================
                 coef    std err          t      P>|t|      [0.025      0.975]
------------------------------------------------------------------------------
const          2.1368      0.015    139.864      0.000       2.107       2.167
dci            0.0016      0.000      4.484      0.000       0.001       0.002
==============================================================================
==============================================================================
                 coef    std err          t      P>|t|      [0.025      0.975]
------------------------------------------------------------------------------
const          2.2060      0.022     99.468      0.000       2.163       2.249
dci            0.0017      0.000      4.655      0.000       0.001       0.002
population    -0.0022      0.001     -4.304      0.000      -0.003      -0.001
==============================================================================
==============================================================================
                 coef    std err          t      P>|t|      [0.025      0.975]
------------------------------------------------------------------------------
const          1.3427      0.027     49.332      0.000       1.289       1.396
dci            0.0022      0.000      6.329      0.000       0.002       0.003
population    -0.0021      0.000     -4.183      0.000      -0.003      -0.001
ln_area        0.3468      0.007     52.308      0.000       0.334       0.360
==============================================================================
```
Interpretation: when considering only users with nonzero boundary distance, an increase of 1 in DCI is associated with a 0.22% change in boundary distance, an increase of 1000 population is associated with a -0.21% change in boundary distance, and a 1% increase in area is associated with a 0.35% change in boundary distance.

**Table 8. Regressions for logarithm of centroid distance**
```
==============================================================================
                 coef    std err          t      P>|t|      [0.025      0.975]
------------------------------------------------------------------------------
const          1.8477      0.016    116.110      0.000       1.817       1.879
dci            0.0014      0.000      3.897      0.000       0.001       0.002
==============================================================================
==============================================================================
                 coef    std err          t      P>|t|      [0.025      0.975]
------------------------------------------------------------------------------
const          2.3612      0.023    101.223      0.000       2.315       2.407
dci            0.0016      0.000      4.436      0.000       0.001       0.002
population    -0.0154      0.001    -29.941      0.000      -0.016      -0.014
==============================================================================
```

```
==============================================================================
                 coef    std err          t      P>|t|      [0.025      0.975]
------------------------------------------------------------------------------
const          1.9963      0.029     68.580      0.000       1.939       2.053
dci            0.0016      0.000      4.479      0.000       0.001       0.002
population    -0.0153      0.001    -29.681      0.000      -0.016      -0.014
ln_area        0.1426      0.007     20.835      0.000       0.129       0.156
==============================================================================
```

Interpretation: when considering all users, an increase of 1 in DCI is associated with a 0.16% change in centroid distance, an increase of 1000 population is associated with a -1.5% change in centroid distance, and a 1% increase in area is associated with a 0.14% change in centroid distance.

**Table 9. Regressions for logarithm of centroid distance, excluding zeros**
```
==============================================================================
                 coef    std err          t      P>|t|      [0.025      0.975]
------------------------------------------------------------------------------
const          2.8395      0.012    235.192      0.000       2.816       2.863
dci            0.0022      0.000      7.923      0.000       0.002       0.003
==============================================================================

==============================================================================
                 coef    std err          t      P>|t|      [0.025      0.975]
------------------------------------------------------------------------------
const          2.9526      0.017    169.603      0.000       2.918       2.987
dci            0.0023      0.000      8.218      0.000       0.002       0.003
population    -0.0036      0.000     -9.009      0.000      -0.004      -0.003
==============================================================================

==============================================================================
                 coef    std err          t      P>|t|      [0.025      0.975]
------------------------------------------------------------------------------
const          2.0253      0.021     96.382      0.000       1.984       2.067
dci            0.0027      0.000      9.924      0.000       0.002       0.003
population    -0.0031      0.000     -8.081      0.000      -0.004      -0.002
ln_area        0.3678      0.005     72.358      0.000       0.358       0.378
==============================================================================
```

Interpretation: when considering only users with nonzero centroid distance, an increase of 1 in DCI is associated with a 0.27% change in centroid distance, an increase of 1000 population is associated with a -0.31% change in centroid distance, and a 1% increase in area is associated with a 0.37% change in centroid distance.

## 11.3.2. Probability of MaxMind matching ground-truth, by user

To regress match probability (Table 10), I use a dependent variable set to 1 when the MaxMind and ground-truth ZIP Code were identical, and 0 otherwise.

**Table 10. Regressions for ground-truth ZIP Code matching MaxMind ZIP Code**

```
==============================================================================
                 coef    std err          t      P>|t|      [0.025      0.975]
------------------------------------------------------------------------------
const          0.1841      0.002     79.522      0.000       0.180       0.189
dci          6.07e-05   5.39e-05      1.126      0.260    -4.5e-05       0.000
==============================================================================
==============================================================================
                 coef    std err          t      P>|t|      [0.025      0.975]
------------------------------------------------------------------------------
const          0.1079      0.003     31.734      0.000       0.101       0.115
dci         3.437e-05   5.36e-05      0.642      0.521   -7.06e-05       0.000
population     0.0023    7.5e-05     30.474      0.000       0.002       0.002
==============================================================================
==============================================================================
                 coef    std err          t      P>|t|      [0.025      0.975]
------------------------------------------------------------------------------
const          0.0402      0.004      9.515      0.000       0.032       0.049
dci         3.737e-05   5.33e-05      0.701      0.483   -6.71e-05       0.000
population     0.0023   7.47e-05     30.980      0.000       0.002       0.002
ln_area        0.0265      0.001     26.642      0.000       0.025       0.028
==============================================================================
```

Interpretation: An increase of 1 in DCI was associated with a 0.00 percentage point change in the probability of a match, an increase of 1000 population was associated with a 0.23 percentage point change, and a 1% increase in area was associated with a 0.027 percentage point change.

## 11.3.3. Number of identified users, by ZIP Code

The dependent variable was the number of users in each ZIP Code according to ground-truth (Table 11) and to MaxMind (Table 12).

**Table 11. Regressions for number of ground-truth users**
```
==============================================================================
                 coef    std err          t      P>|t|      [0.025      0.975]
------------------------------------------------------------------------------
const          7.9334      0.129     61.331      0.000       7.680       8.187
dci           -0.0612      0.002    -24.734      0.000      -0.066      -0.056
==============================================================================
==============================================================================
                 coef    std err          t      P>|t|      [0.025      0.975]
------------------------------------------------------------------------------
const          2.6124      0.137     19.039      0.000       2.343       2.881
dci           -0.0556      0.002    -25.855      0.000      -0.060      -0.051
population     0.2517      0.004     67.500      0.000       0.244       0.259
==============================================================================
==============================================================================
                 coef    std err          t      P>|t|      [0.025      0.975]
------------------------------------------------------------------------------
const          6.1007      0.184     33.089      0.000       5.739       6.462
dci           -0.0472      0.002    -22.284      0.000      -0.051      -0.043
population     0.2389      0.004     65.234      0.000       0.232       0.246
ln_area       -1.1184      0.041    -27.464      0.000      -1.198      -1.039
==============================================================================
```

Interpretation: an increase of 1 in DCI was associated with 0.047 fewer users according to ground-truth, an increase of 1000 population is associated with 0.24 more users, and an increase of 1% in area is associated with 0.012 fewer users.

**Table 12. Regressions for number of MaxMind users**

```
==============================================================================
                 coef    std err          t      P>|t|      [0.025      0.975]
------------------------------------------------------------------------------
const          7.2471      0.153     47.512      0.000       6.948       7.546
dci           -0.0497      0.003    -17.039      0.000      -0.055      -0.044
==============================================================================

==============================================================================
                 coef    std err          t      P>|t|      [0.025      0.975]
------------------------------------------------------------------------------
const          1.4310      0.165      8.648      0.000       1.107       1.755
dci           -0.0436      0.003    -16.807      0.000      -0.049      -0.039
population     0.2751      0.004     61.177      0.000       0.266       0.284
==============================================================================

==============================================================================
                 coef    std err          t      P>|t|      [0.025      0.975]
------------------------------------------------------------------------------
const          4.9947      0.224     22.293      0.000       4.556       5.434
dci           -0.0350      0.003    -13.598      0.000      -0.040      -0.030
population     0.2620      0.004     58.883      0.000       0.253       0.271
ln_area       -1.1426      0.049    -23.089      0.000      -1.240      -1.046
==============================================================================
```

Interpretation: an increase of 1 in DCI was associated with 0.035 fewer users according to ground-truth, an increase of 1000 population is associated with 0.26 more users, and an increase of 1% in area is associated with 0.011 fewer users.